\documentclass[12pt,a4paper]{article}

\usepackage{amsfonts,amssymb,amsmath,amsopn,amsthm,graphicx}

 \numberwithin{equation}{section}

\newtheorem{thm}{Theorem}[section]
\newtheorem{defin}{Definition}[section]
\newtheorem{lem}{Lemma}[section]
\newtheorem{cor}{Corollary}[section]
\newtheorem{prop}{Proposition}[section]
\newtheorem{rem}{Remark}[section]

\newcommand{\be}{\begin{equation}}
\newcommand{\ee}{\end{equation}}
\newcommand{\bea}{\begin{eqnarray}}
\newcommand{\eea}{\end{eqnarray}}

\newcommand{\pd}{\partial}
\newcommand{\D}{\,\textrm{d}}
\newcommand{\Le}{\left}
\newcommand{\Ri}{\right}

\newcommand{\N}{\mathbb{N}}

\newcommand{\R}{\mathbb{R}}
\newcommand{\C}{\mathbb{C}}

\newcommand{\ra}{\rightarrow}

 \DeclareMathOperator{\Int}{Int}
 
 \DeclareMathOperator{\supp}{supp}
\DeclareMathOperator{\dist}{dist} \DeclareMathOperator{\Ran}{Ran}

\begin{document}

\title{Resonances Width in Crossed Electric and Magnetic Fields}
\author{Christian Ferrari$^a$ and Hynek Kova\v r\'{\i}k$^{b}$}
\date{}
\maketitle

\begin{flushleft}
 {\em a) Institute for Theoretical Physics, Ecole
 Polytechnique F\'ed\'erale de Lausanne, CH-1015  Lausanne, Switzerland
 \\ b) Institut f\"ur Analysis, Dynamik und Modellierung, Universit\"at
 Stuttgart, Pfaffenwaldring 57, D-70569 Stuttgart, Germany}\footnote[1]{also
    on leave from  Department of Theoretical Physics, Nuclear Physics
    Institute, Academy of Sciences, 25068 \v{R}e\v{z} near Prague, Czech Republic}
\end{flushleft}

\begin{abstract}
We study the spectral properties of a charged particle confined to
a two-dimensional plane and submitted to homogeneous magnetic and
electric fields and an impurity potential $V$. We use the method
of complex translations to prove that the life-times of resonances
induced by the presence of electric field are at least Gaussian
long as the electric field tends to zero.
\end{abstract}

\section{Introduction}

\noindent The purpose of this paper is to study the dynamics of an
electron in two dimensions in the presence of crossed magnetic and
electric fields and a potential type perturbation. We assume that
the magnetic field acts in the direction perpendicular to the
electron plane with a constant intensity $B$ and that the electric
field of constant intensity $F$ points in the $x-$direction. The
perturbation $V(x,y)$ is supposed to satisfy certain localisation
conditions. The corresponding quantum Hamiltonian reads as follows
$$
 H(F) = H(0)-Fx=H_L+V-Fx,
$$
where $H_L$ is the Landau Hamiltonian of an electron in a
homogeneous magnetic field of intensity $B$. Its spectrum is given
by the infinitely degenerate eigenvalues (Landau levels)
$(2n+1)B$, $n\in \N$.
\par
When $F=0$, the impurity potential $V$ creates generically an
infinite number of eigenvalues of $H(0)$ in between the Landau
levels. These eigenvalues, which correspond to the so-called impurity states,
then accumulate at Landau levels. This holds for
any sign definite, bounded $V$, which tends to zero at infinity,
see \cite{raik}, \cite{MR}. Classically, such impurity states
represent the electron motion on localised trajectories. The main
question that we address is what happens with these localised
states when a constant electric field is switched on. In
particular one would like to know, whether the eigenvalues of
$H(0)$ may survive in the presence of a nonzero electric field and
if not, what is the characteristic time in which they dissolve.
\par
Answer to this question is well known for the hydrogen atom in a
homogeneous electric field, in which case the corresponding
Schr\"odinger operator has no eigenvalues, \cite{tit}. The
localised states turn into so-called Stark resonances, whose
life-times are exponentially long as $F\ra 0$. This was first
computed by Oppenheimer in \cite{oppen} and later rigorously
proved in \cite{HSimon}. The  Oppenheimer formula was then
partially generalised also for many body and non Coulombic
potentials, see \cite{sig} and references therein.
\par
On the other hand, results concerning systems with simultaneous
constant magnetic and electric fields are scarce. Such a model is
considered in \cite{GM} where the impurity $V$ is supposed to act
as a $\delta-$potential. Using the special properties of a
two-dimensional $\delta-$interaction, the authors of \cite{GM}
compute the spectral density of $H(F)$ in the neighbourhood of the
discrete spectrum of $H(0)$ and prove that all impurity states are
unstable. Their life-times are then shown to be of order $\exp[
\frac{B}{F^2}]$ as $F\ra 0$ and it is conjectured that such a
behaviour holds in general. It is our motivation to extend this
result  for continuous impurity potentials when the method of
\cite{GM} is no longer applicable. In particular, we will prove
under some assumptions on $V$ that the life-times of magnetic
Stark resonances are for $F$ small enough at least Gaussian long,
i.e. we find a lower bound compatible with the asymptotics
obtained in \cite{GM}.
\par
Let us now describe the content of our paper more in detail. Basic
mathematical tool we use is the method of complex translations for
Stark Hamiltonians, which was introduced in \cite{AH} as a
modification of the original theory of complex scaling \cite{AC},
\cite{BC}. Following \cite{AH} we consider the transformation
$U(\theta)$, which acts as a translation in $x-$direction;
$(U(\theta)\psi)(x)=\psi(x+\theta)$. For non real $\theta$ the
translated operator $H(F,\theta)=U(\theta)H(F) U^{-1}(\theta)$ is
non-selfadjoint and therefore can have some complex eigenvalues.
The complex eigenvalues of $H(F,\theta)$ with $\Im \theta>0$ are called
the spectral resonances of $H(F)$, see e.g. \cite{HSigal}, and the corresponding 
resonance widths are given by their imaginary parts.
Moreover, the
result of \cite{FK1} tells us that if $\phi$ is an
eigenfunction of $H(0)$, then $(\phi,e^{-itH(F)}\, \phi)$ decays
exponentially at the rate given by the imaginary parts of the
eigenvalues of $H(F,\theta)$.
\par
In Section \ref{location2} we show that the eigenvalues of
$H(F,\theta)$ are located in the Gaussian small vicinity of real
axis as $F\ra 0$, see Theorem \ref{main}. In order to prove this
we employ a geometric resolvent equation in the form developed in
\cite{BG} for the study of Stark Wannier Ladders. The idea of our
proof is based on the fact that the eigenfunctions of  $H(0)$ have
a Gaussian-like decay at infinity and therefore ``feel'' the
electric field only locally. That leads us to a construction of
the reference Hamiltonian $H_2(F)$, which describes the system
where the electric field is localised in the vicinity of impurity
potential $V$ by a suitable cut-off function. For a precise
definition of $H_2(F)$ see Section \ref{location1}. When $F\ra 0$
we let the cut-off function tend to $1$ at the rate proportional
to $F^{-1+\varepsilon}$ ($\varepsilon>0$), which assures the
convergence of spectra of $H_2(F)$ to that of $H(0)$. It follows
from the general theory of complex deformations that the discrete
spectrum of $H_2(F)$ is not affected by the transformation
$U(\theta)$. Moreover, for $H_2(F)$ also the essential spectrum
does not change under $U(\theta)$. Therefore
$\sigma(H_2(F,\theta))$ remains real even when $\theta$ becomes
complex. The geometric resolvent equation, (\ref{decoupling}),
then allows us to deduce that for $F$ small enough the resolvent
$R(z;\theta)=(z-H(F,\theta))^{-1}$ is bounded except in a small
neighbourhood of the eigenvalues of $H_2(F,\theta)$. More
precisely, we show that the norm of $R(z;\theta)$ remains bounded
as long as the distance between $z$ and $\sigma(H_2(F,\theta))$ is
at least of order \be \label{asymptotics} e^{- \frac{B\,
C}{F^{2(1-\varepsilon)}}},\quad \varepsilon>0, \ee where $C$ is a
strictly positive constant and $\varepsilon$ can be taken
arbitrarily small. Moreover, we prove that on the energy intervals
well separated from Landau levels the spectral projector of
$H(F,\theta)$ converges uniformly to that of $H_2(F,\theta)$ as
$F\ra 0$. These results give us the existence of eigenvalues of
$H(F,\theta)$ and an upper bound on their imaginary parts. Let us
note, that our result does not exclude the existence of point
spectrum of $H(F)$. In other words, we do not answer the question
whether all impurity states become unstable once the electric
field with finite intensity is switched on. Although the quantum
tunnelling phenomenon leads us to believe that it is indeed the
case, a rigorous proof is missing and the question remains open.
\par

\section{The Model}\label{modelres2}

We work in the system of units, where $m=1/2,\, e=1,\, \hbar =1$.
The crossed fields Hamiltonian is then given by \be H_1(F)= H_L-Fx
= (-i\pd_x+By)^2 -\pd_y^2-Fx,\quad {\rm on}\quad L^2(\R^2). \ee
Here we use the Landau gauge with ${\bf A}(x,y)=(-By,0)$. A
straightforward application of \cite[Thm. X.37]{rees} shows that
$H_1(F)$ is essentially self-adjoint on $C_0^{\infty}(\R^2)$, see
also \cite[Prob. X.38]{rees}. Moreover, one can easily check that
\begin{equation}\label{specH1nd}
\sigma(H_1(F))=\sigma_{ac}(H_1(F))=\R
\end{equation}
As mentioned in the Introduction we employ the translational
analytic method developed in \cite{AH}. We introduce the
translated operator $H_1(F,\theta)$ as follows:
\begin{equation}
H_1(F,\theta)= U(\theta)H_1(F)U^{-1}(\theta)
\end{equation}
where
\begin{equation}
\left(U(\theta)f\right)(x,y):=\left(e^{ip_x\theta}f\right)(x,y)=f(x+\theta,
y)
\end{equation}
An elementary calculation shows that
\begin{equation}
H_1(F,\theta)=H_1(F)-F\theta
\end{equation}
Operator $H_1(F,\theta)$ is clearly analytic in $\theta$.
Following \cite{AH} we define the class of $H_1(F)-$translation
analytic potentials.

\begin{defin}
Suppose that $V(z,y)$ is analytic in the strip $|\Im z|<\beta$,
$\beta>0$ independent of $y$. We then say that $V$ is
$H_1(F)-$translation analytic if $V(x+z,y)(H_1(F)+i)^{-1}$ is a
compact analytic operator valued function of $z$ in the given
strip.
\end{defin}

We can thus formulate the conditions to be imposed on $V$:
\begin{itemize}
\item [$(a)$] $V(x,y)$ is $H_1(F)-$translation analytic in the
strip $|\Im z|<\beta$. \item[(\emph{b})] There exists $\beta_0\leq
\beta$ such that for $|\Im z|\leq \beta_0$ the function $V(x+z,y)$
satisfies
$$ |V(x+z,y)|\leq
\begin{cases} V_0 & \textrm{ if } x \in [-a_0-\Re z,a_0-\Re z], y \in [-a_1,a_1] \\
V_0\, e^{-\nu\, (x+\Re z)^2},\, \nu>0 & \textrm{ if } x\not \in
[-a_0-\Re z,a_0-\Re z]
\end{cases}
$$
and
$$ |V(x+z,y)|=0,\quad y\not\in[-a_1,a_1]
$$
\end{itemize}
for given positive constants $a_0,a_1$, independent of $F$.

In order to characterise the potential class for which the above
conditions are fulfilled let us assume for the moment, that the
integral kernel of $(H_1(F)+i)^{-1}$ has at most a local
logarithmic singularity at the origin. This is a very plausible
hypothesis, see Lemma 4.3 in \cite{FK1}, it then follows that any
$L^2(\R^2)$ function that can be analytically continued in a given
strip $|\Im z|<\beta$ satisfies the condition $(a)$. If in
addition the analytic continuation satisfy $(b)$, both assumptions
are satisfied.

\begin{rem} It follows from the proof of our main result, given below,
  that the
  localisation of $V$ w.r.t. $y$ could be replaced by a Gaussian
  decay. However, we use the assumption $(b)$ in order to keep
  the computations as simple as possible. Note that this assumption is of
  crucial importance to get the Gaussian upper bound, in $1/F$, on the imaginary
  part of the eigenvalues of $H(F,ib)$. See in particular Remark
  \ref{condonV} in Appendix A.
  \end{rem}

From the well known perturbation argument, \cite{Kato}, we see
that under assumption $(b)$
\begin{equation}
H(F,\theta) = U(\theta)H(F)\, U^{-1}(\theta)= H_1(F,\theta)+V\left
(x+\theta,y\right )
\end{equation}
forms an analytic family of type $A$. \par Furthermore, since
$V(x+\theta,y)(H_1(F)+i)^{-1}$ is compact by $(a)$, we have
\cite[Cor. 2, p. 113]{rees}
\begin{equation}\label{specH1a}
\sigma_{ess}(H(F,\theta)+ibF) = \sigma_{ess}(H_1(F)) = \R \;
\;\Longrightarrow\;\; \sigma_{ess}(H(F,\theta))= \R -ibF
\end{equation}where $\theta=i b,\, b\in\R$. By standard arguments \cite[Prob.
XIII.76]{rees}, all eigenvalues of $H(F,ib)$ lie in the strip
$-bF<\Im z\leq 0$ and are independent of $b$ as long as they are
not covered by the essential spectrum. \par
\noindent The complex eigenvalues of $H(F,\theta)$ with $\Im\theta>0$, in
$\{z\in \C:-bF<\Im z<0\}$ are called the spectral {\it
resonances} of $H(F)$, and are intrinsic to $H(F)$, see \cite[Chap.\ 16]{HSigal}.
The corresponding \emph{resonance widths} are given by the 
imaginary parts of the eigenvalues $E_\alpha$ of $H(F,\theta)$: $\Gamma_\alpha=-2\Im
E_\alpha$, and the \emph{lifetimes} by $\tau_\alpha=\Gamma_\alpha^{-1}$. \\

Next we will show that, for sufficiently weak electric field $F$,
the eigenvalues $E_\alpha$ of $H(F,ib)$ exist and are located in
Gaussian small neighbourhood of real axis. In particular, we will
prove that
$$
|\Im E_\alpha|\leq e^{-
\frac{B\tilde{R}_\alpha}{F^{2(1-\varepsilon)}}}
$$
where the positive constant $\tilde{R}_\alpha$ depends on the real
part of $E_\alpha$ and $\varepsilon$ can be made arbitrarily
small. The method we employ is based on the decoupling formula
developed in \cite{BG}, see also \cite{FM1}.

\section{Auxiliary Hamiltonian}\label{location1}

\noindent The reference Hamiltonian reads
$$
 H_2(F)= H_L+V-Fx h_F(x)\chi_A(y)\equiv H_L+V+W_F
$$
with $\chi_A$ being characteristic function of the set
$A=[-\bar{y},\bar{y}]$ ($\bar{y}=y_1+\frac{1}{F^\tau}$, with $y_1$
and $\tau$ defined in Section \ref{decsch} below) and
$$
h_F(x)=\tfrac{1}{2}
\left\{\tanh(\gamma_F(x+\bar{x}))-\tanh(\gamma_F(x-\bar{x}))\right\}
$$
where\footnote{We will often drop the subscript $F$.}
$\gamma_F=\frac{\gamma_0}{F^{1-\varepsilon}}>0$ and $\bar{x}>0$
must satisfy
\begin{equation}\label{VCbar}
F\bar{x}\to 0 \quad \text{as}\quad F\to 0\; .
\end{equation}

This is required because we don't want the local electric field
to modify significatively the impurity potential $V$. We 
can thus expect that the spectrum of $H_2(F)$ is ``close'' to
that of $H(0)$. We will chose $\bar{x}=
\frac{\bar{C}}{F^{1-\varepsilon}}>0$, for $\varepsilon>0$.\\ In
Figure 1 we sketch the $x-$section of $V(x,y)-xh_F(x)\chi_A(y)$
for the case of impurity potential given by
$V(x,y)=-V_0e^{-x^2}f(y)$ ($f$ being any locally supported
positive bounded function).

\begin{figure}[!h]\label{pot1}
\begin{center}
\input{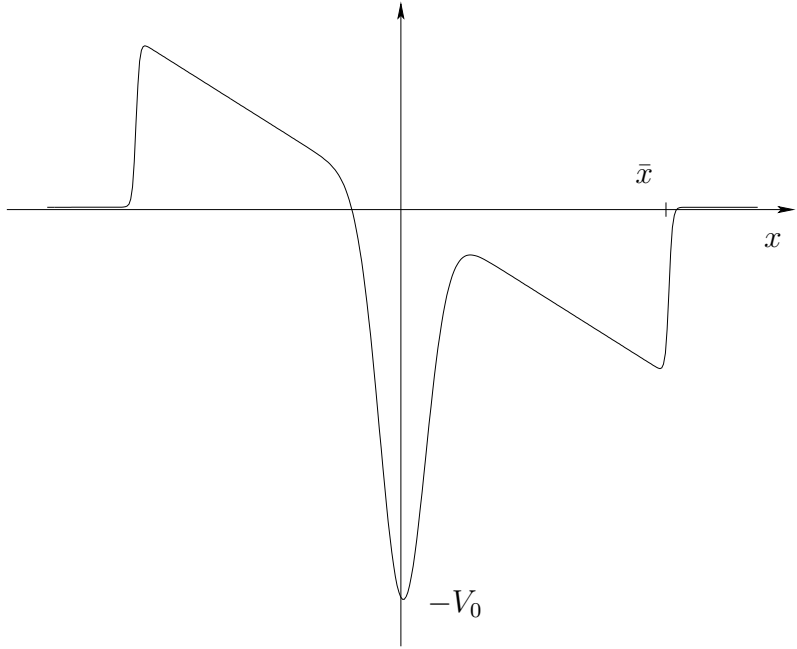}
\end{center}
\caption{\emph{The $x-$section for the potential of $H_2(F)$
satisfying condition \eqref{VCbar} for a negative Gaussian
potential.}}
\end{figure}
Before giving the results on the spectral properties of $H_2(F)$
and its translated correspondent $H_2(F,ib)$ we define the set of
$\theta=ib$ for which $W_F$ can be analytically continued in the
$x$ variable. Since $\tanh(z)$ has an analytic continuation for
$|\Im z|<\frac{\pi}{2}$ we have $\gamma_F |b|<\frac{\pi}{2}$. For
our purpose we will consider the family of operator
$U(\theta)\equiv U(ib)$ defined in Section 1, with $\theta\in
{\cal D}_\theta$ where
$$
{\cal D}_\theta=\{\theta\in \C: \gamma_F|\Im
\theta|<\tfrac{\pi}{4}\}
$$
Since $\gamma_F=\frac{\gamma_0}{F^{1-\varepsilon}}$ we take
\begin{equation}
\label{bsmall} b=b_0 F^{\alpha}, \alpha>2
\end{equation}

\begin{prop}$ $\label{prop21}
Assume $V$ satisfies $(a)$ and $(b)$. Then
\begin{enumerate}
\item For each $e_\alpha\in \sigma(H(0))$ there is a family of
$\lambda_\alpha(F)\in \sigma(H_2(F))$ such that
$\lambda_\alpha(F)\to e_\alpha$ for $F\to 0$. \item Let
$P_\Delta(F)$ respectively $P_\Delta(0)$ be the eigenprojector of
$H_2(F)$ respectively $H(0)$ on the interval $\Delta$. Then
$\|P_\Delta(F)-P_\Delta(0)\|\to 0$ as $F\to 0$. \item $
\sigma_{ess}(H_2(F)) = \sigma_{ess}(H_L)= \{(2n+1)B; n\in \N\} $
\item For each $e_\alpha\in \sigma_d(H(0))$ there exists a
constant $c$ such that
\begin{equation*}
\lambda_\alpha(F)\in
[e_\alpha-cF^\varepsilon,e_\alpha+cF^\varepsilon]
\end{equation*}
\end{enumerate}
\end{prop}

\begin{proof}
We have
\begin{eqnarray}
\|(H(0)-z)^{-1} - (H_2-z)^{-1}
\|&=&\|(H_2-z)^{-1}[H_2-H(0)](H(0)-z)^{-1}\| \nonumber\\
&\leq& \|(H_2-z)^{-1}\|\|(H_2-H(0))\|\|(H(0)-z)^{-1}\| \nonumber\\
&\le& \frac{1}{|\Im z|^2}\|Fxh_F(x) \chi_A(y)\| \rightarrow 0
\end{eqnarray}
as $F\to 0$ due to the choice of $h_F$. Thus $H_2(F)\to H(0)$ in
the norm resolvent sense. The Statement 1. and 2. of the Lemma now
follows from \cite[Thm. VIII.1.14]{Kato} and \cite[Thm.
VIII.23]{rees}. Statement 3. follows from  the fact that $W_F$ and
$V$ are $H_L-$compact, see proof of Lemma \ref{spectofh2} below.
Finally the estimate
\begin{eqnarray}
\|Fxh_F(x) \chi_A(y)\| \leq F \|xh_F(x)\|_\infty \leq
cF^{\varepsilon}
\end{eqnarray}
yields Statement 4.
\end{proof}

We now show that the spectrum of $H_2(F)$ is not affected by the
transformation $U(ib)$:

\begin{lem} \label{spectofh2} Under the assumptions of Proposition
\ref{prop21} $\{H_2(F,\theta):\, \theta\in {\cal D}_\theta\}$
forms a self-adjoint holomorphic family of type $A$. Moreover, for
each $ib \in {\cal D}_\theta$ one has
\begin{eqnarray*} \sigma_{ess}(H_2(F,ib))&=& \sigma_{ess}(H_2(F))  \\
\sigma_d(H_2(F,ib))&=& \sigma_d(H_2(F))
\end{eqnarray*}
\end{lem}

\begin{proof}
To prove that $\{H_2(F,\theta):\, \theta\in {\cal D}_\theta\}$
forms a self-adjoint holomorphic family we have show that
$H_2(F,\theta)$ is holomorphic w.r.t. $\theta\in {\cal D}_\theta$
and that its domain is independent of $\theta$, see \cite[pp. 375,
385]{Kato}. First claim follows from the assumptions on $V$ and
from the explicit form of $W_F$. The boundedness of $V,\, W_F$
then implies the $\theta-$independence of the domain. For the the
stability of essential spectrum we recall \cite[Thm.
18.8]{HSigal}, which tells us that it is enough to prove that
$W_F(x+ib,y)(H_L+i)^{-1}$ and $V(x+ib,y)(H_L+i)^{-1}$ are compact.
We first observe that
$$
h_F(x+ib)=\frac{e^{2\gamma_F\bar{x}}-e^{-
2\gamma_F\bar{x}}}{e^{2\gamma_F\bar{x}}+e^{-2\gamma_F\bar{x}}+
e^{2\gamma_F(x+ib)}+e^{-2\gamma_F(x+ib)}}.
$$
Thus
$$
|h_F(x+ib)|\leq\frac{e^{2\gamma_F\bar{x}}}{\left[e^{2\gamma_F
x}+e^{-2\gamma_F x}\right]\cos(2\gamma_F
b)+\left[e^{2\gamma_F\bar{x}}+e^{-2\gamma_F\bar{x}}\right]}
$$
From the latter estimate we deduce that $\lim_{x\to \pm
\infty}|W_F(x+ib,y)|=0$ and that $|W_F(x+ib,y)|$ is uniformly
bounded. Since $\chi_A$ has compact support, $W_F(ib)\in
L^2(\R^2)$. Then
\begin{eqnarray}
\|W_F(ib)(H_L+i)^{-1}\|_{HS}^2 &=& \int_{\R^2} \D {\bf x}
|W_F(x+ib,y)|^2 \int_{\R^2} \D {\bf x}' |G_L({\bf x},{\bf
x}';i)|^2
\nonumber\\
&=&\int_{\R^2} \D {\bf x} |W_F(x+ib,y)|^2 \int_{\R^2} \D {\bf u}
|G_L({\bf u};i)|^2 < \infty
\end{eqnarray}
where $|G_L({\bf x},{\bf x}';i)|=|G_L({\bf x}-{\bf
x}';i)|=|G_L({\bf u};i)| \in L^2(\R^2)$ is the integral kernel of
$(H_L+i)^{-1}$, see for example \cite{CN}. Hence
$W_F(ib)(H_L+i)^{-1}$ is compact. Same argument shows that also
$V(ib)(H_L+i)^{-1}$ is compact.

Finally the stability of the discrete spectrum follows from a
standard analyticity argument \cite[Prob. XIII.76]{rees}.
\end{proof}

\noindent We now give a result on the norm of $R_2(z;ib)$, which
 will be used later in the proof of our main theorem. \footnote{ Henceforth
the symbol $\cal C$ denotes a strictly positive real number
independent of $F$.}

\begin{lem}\label{R_2}
Let $z \in \C$ such that $(2q-1)B+\delta<\Re z<(2q+1)B-\delta$
$(\delta>0)$ for some $q\in \N$. Then there exists a natural
number $0<s<\infty$, such that
$$
\|R_2(z;ib)\|\leq {\cal C} \,|\Im z|^{-s},
$$
holds true provided $F$ is small enough.
\end{lem}

\begin{proof}
We introduce the operator $A(ib)$ by
\begin{equation}\label{Aib}
A(ib)=H_2(ib)-H_2
\end{equation}
(here we note $H_2(ib)\equiv H_2(F,ib)$ and $H_2\equiv H_2(F)$).
>From the definition of $H_2(ib)$ it easily follows that there
exists certain constant $A_0$ such that for $b=b_0F^{\alpha}$
$$ \|A(ib)\|\leq A_0
F^{\alpha-1+\varepsilon}(1+\mathcal{O}(F^{\alpha}))
$$
We need a preliminary result. A standard perturbation argument now
shows that if
$$
{\rm dist}\, (\sigma(H_2(F)),\xi)= d_0 F^\varepsilon
$$
then
\begin{equation} \label{normr2} \|R_2(\xi;ib)\|\leq
\frac{\|R_2(\xi;0)\|}{1-\|A(ib)R_2(\xi;0)\|}=F^{-\varepsilon}\,
\frac{1}{d_0-F^{\alpha-1}A_0}
\end{equation}
whenever $d_0>F^{\alpha-1}A_0$, i.e. whenever $F$ is small enough.
To continue let $e_\alpha$ be the eigenvalue of $H(0)$ which
minimises $| z-(e_\alpha\pm cF^\varepsilon)|$. We define a circle
$\tilde{\Gamma}\equiv\{\xi\in
\C:|\xi-e_\alpha|=\Gamma_0F^{\varepsilon}\}$ enclosing only the
eigenvalues of $H_2(F)$ converging to $e_\alpha$ for given
$e_\alpha$. Let $P^{\tilde{\Gamma}}_2(ib)$ the projector onto
$\overline{\Int \tilde{\Gamma}}$ associated to $H_2(ib)$
$$
P^{\tilde{\Gamma}}_2(ib)\equiv P_2(ib)= \frac{1}{2\pi i}
\oint_{\tilde{\Gamma}} R_2(z;ib) \D z
$$
Since $P_2(ib)$ is a projector, applying
\cite[Thm.III.6.17]{Kato}, the resolvent of $H_2(ib)$ decomposes
as follows
$$
R_2(z;ib)=R_2'(z;ib) + R_2''(z;ib)
$$
where
\begin{eqnarray}
R_2'(z;ib)&=&P_2(ib)R_2'(z;ib)=R_2'(z;ib)P_2(ib)\\
R_2''(z;ib)&=&[1-P_2(ib)]R_2'(z;ib)=R_2'(z;ib)[1-P_2(ib)]
\end{eqnarray}
Let $H'$ be the restriction of $H_2(ib)$ on $M'\equiv \Ran
P_2(ib)$ and $H''$ the restriction of $H_2(ib)$ on $M''\equiv \Ran
[1-P_2(ib)]$. From \cite[Thm.III.6.17]{Kato} it follows that
$R_2'(z;ib)$ coincides with $(z-H')^{-1}$ on $M'$ and vanishes on
$M''$. Similarly $R_2''(z;ib)$ coincides with $(z-H'')^{-1}$ on
$M''$ and vanishes on $M'$. Since $\dist(\sigma(H''),z)$ is bounded
from below by a constant we can use \eqref{normr2} to get
$$
\|R_2''(z;ib)\|\leq {\cal C}
$$
Let us denote $r_0=\dim P_2(ib)$. We can then write
$$
R_2'(z;ib)=\sum_{h=1}^{r_0} \left[(z-\zeta_h)^{-1}\,
P_h+(z-\zeta_h)^{-1}\sum_{n=1}^{m_h-1}
  (z-\zeta_h)^{-n}D_h^n\right]
$$
where $\zeta_h\equiv \lambda_{\alpha,h}\in\R$ are the eigenvalues
of $H'$, $P_h$ the corresponding projectors, $m_h=\dim P_h$ and
$D_h$ denotes the nilpotent associated to $\zeta_h$, see
\cite[Chap.I]{Kato}. So we can always find some $s\in\N$ ($1\leq s
\leq \max_h m_h\leq r_0$), such that
$$
\|R_2'(z;ib)\| \leq {\cal C}\, {\rm dist}(z,\sigma(H'))^{-s} \leq
{\cal C}\, |\Im z|^{-s},
$$
which concludes the proof.
\end{proof}

\section{Setup of a decoupling scheme}\label{decsch}

As already mentioned in the Introduction, the eigenfunctions of $H(0)$
 ``feel'' the electric field only
locally and the properties of the Hamiltonian $H(F)$ can be
derived on the basis of those of the ``local field'' Hamiltonian
$H_2(F)$ described above. To make this idea 
work we use the geometric resolvent perturbation theory in the form
developed in \cite{BG} (see also \cite{BCD}, \cite{HSigal}). It
consists of dividing the configuration space $\R^2$ in different 
regions and study of Hamiltonians $H_i$ with associated potentials $V_i$
which are in the considered regions close to that of the full Hamiltonian
 $H(F)$.

We introduce the following functions that give a decoupling along
the $x-$axis.
\begin{eqnarray}
J_{-}(x)&=&\tfrac{1}{2}\left[1+\tanh (\gamma_F(x-x_2)) \right] \nonumber\\
\tilde{J}_{-}(x)&=&\tfrac{1}{2}\left[1+\tanh (\gamma_F(x-x_0))
\right]
\nonumber\\
J_{0}(x)&=&\tfrac{1}{2}\left[\tanh (\gamma_F(x+x_1)) -\tanh
(\gamma_F(x-
x_1))\right] \nonumber\\
\tilde{J}_{0}(x)&=&\tfrac{1}{2}\left[\tanh (\gamma_F(x+x_0))-\tanh
(\gamma_F(x-x_0)) \right] \nonumber\\
J_{+}(x)&=&\tfrac{1}{2}\left[1-\tanh (\gamma_F(x+x_2)) \right] \nonumber\\
\tilde{J}_{+}(x)&=&\tfrac{1}{2}\left[1-\tanh (\gamma_F(x+x_0))
\right]
\end{eqnarray}
where $0<x_2=\frac{C_2}{F^{1-\varepsilon}}<
x_0=\frac{C_0}{F^{1-\varepsilon}} < x_1=\frac{C_1}
{F^{1-\varepsilon}}< \bar{x}$.
 \noindent Along the $y-$axis we use three bounded
$C^\infty(\R)$ functions
\begin{eqnarray}
J_<(y)&=&
\begin{cases}
1 \quad \text{if } y\leq -y_0+\frac{1}{F^{\tau}} \\
0 \quad \text{if } y\geq -y_2
\end{cases} \qquad
J_c(y)=
\begin{cases}
1 \quad \text{if }  |y|\leq y_0+\frac{1}{F^{\tau}} \\
0 \quad \text{if }  |y|\geq y_1
\end{cases}
\nonumber\\
J_>(y)&=&
\begin{cases}
1 \quad \text{if } y\geq  y_0-\frac{1}{F^{\tau}}\\
0 \quad \text{if } y\leq y_2
\end{cases}
\end{eqnarray}
where $0<y_2=a_1+1$, $y_0=y_2+\frac{1}{F^{\tau}}+1$,
$y_1=y_0+\frac{1}{F^{\tau}}+1$, where $\tau>\alpha+2$. We will
also assume that $\|J'_i\|_{\infty},\, \|J''_i\|_{\infty}<\infty$,
$i\in \{<,\,>,\, c\}$.\\
Note that for the $x-$cut the dependence on $F$ of $x_0,x_1,x_2$
is the optimal choice to get the desired results, while in the
$y-$cut the dependence on $F$, i.e. the factor $F^{-\tau}$, is
such that $\tau$ can be chosen as large as we need.

\noindent The system is then cut in five parts according to the
following ``full'' decoupling functions (see Figure 2):
\begin{align}
& \begin{cases}
J_1(x,y)&=J_{-}(x)J_c(y) \\
\tilde{J}_1(x,y)&=\tilde{J}_-(x)\tilde{J}_c(y)
\end{cases} \qquad
&\begin{cases}
J_2(x,y)&=J_0(x)J_c(y) \\
\tilde{J}_2(x,y)&=\tilde{J}_0(x)\tilde{J}_c(y)
\end{cases}
\nonumber\\
& \begin{cases}
J_3(x,y)&=J_>(y) \\
\tilde{J}_3(x,y)&=\tilde{J}_>(y)
\end{cases} \qquad
& \begin{cases}
J_4(x,y)&=J_<(y) \\
\tilde{J}_4(x,y)&=\tilde{J}_<(y)
\end{cases}
\qquad
\begin{cases}
J_5(x,y)&=J_{+}(x)J_c(y) \\
\tilde{J}_5(x,y)&=\tilde{J}_+(x)\tilde{J}_c(y)
\end{cases}
\nonumber
\end{align}
with
$$
\tilde{J}_<(y)=\chi_{(-\infty,-y_0]}(y), \quad
\tilde{J}_c(y)=\chi_{[-y_0,y_0]}(y),\quad
\tilde{J}_>(y)=\chi_{[y_0,\infty)}(y)
$$

\begin{figure}[!h]\label{dec2D}
\begin{center}
\input{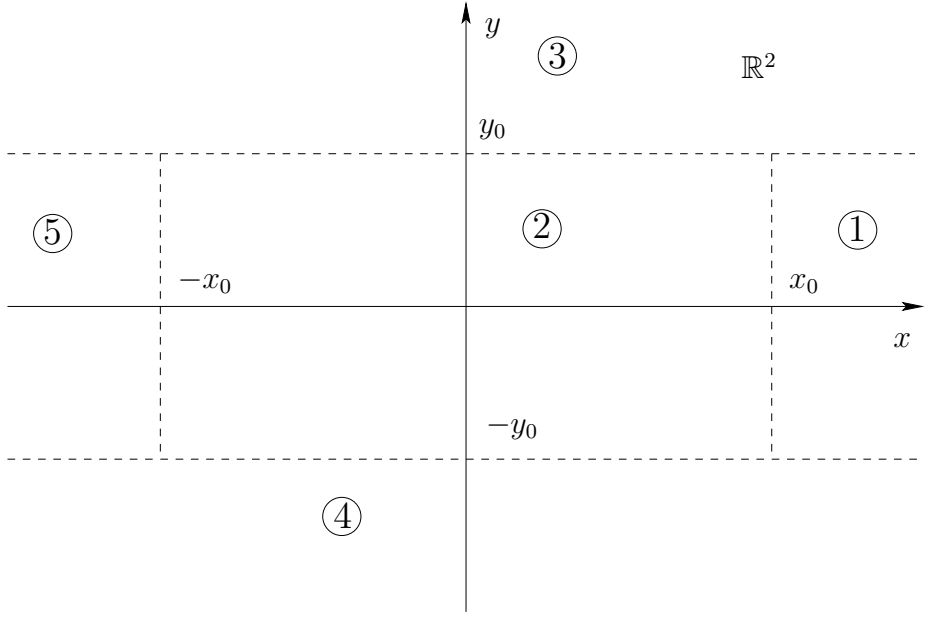}
\end{center}
\caption{\emph{Schematic representation of the decoupling scheme.
In region 2 the total potential $V(x,y)-Fx$ is close to the local
potential of the auxiliary Hamiltonian $H_2(F)$, while in the
others it is close to the electric potential $-Fx$.}}
\end{figure}

We remark that all these functions have an analytic continuation
in the $x$ variable ($x\to x+ib$) if $ib\in {\cal D}_\theta$.

We are now ready to establish the decoupling scheme. We introduce
the following auxiliary Hamiltonians: $H_3=H_4=H_5=H_1=H_L-Fx$ and
$H_2(F)\equiv H_2$ treated in the previous paragraph. For
simplicity we write $H$ for $H(F)$.

\noindent Note that
$$
HJ_1=H_1J_1+VJ_1, \quad HJ_5=H_5J_5+VJ_5, \quad HJ_3=H_3J_3, \quad
HJ_4=H_4J_4
$$
and, using $\chi_A(y)J_c(y)=J_c(y)$,
$$
HJ_2=H_2J_2 -Fx(1-h_F)(x)J_2
$$
thus
\begin{equation}\label{bga}
(z-H)\sum_{i=1}^5 J_i R_i(z) \tilde{J}_i =\sum_{i=1}^5(z-H_i)J_i
R_i(z) \tilde{J}_i + A_1+A_5 +A_2 =1- K(z)
\end{equation}
where $A_1=VJ_1R_1(z)\tilde{J}_1$, $A_5=VJ_5R_5(z)\tilde{J}_5$,
$A_2=-Fx(1-h_F)(x)J_2R_2(z)\tilde{J}_2$ and
$$
 K(z)=\sum_{i=1}^5[H_L,J_i]R_i(z)\tilde{J}_i + \left(\sum_{i=1}^5
J_i\tilde{J}_i-1\right) - A_1 -A_5 - A_2
$$
From \eqref{bga} we deduce the decoupling formula
\begin{equation}\label{res}
R(z)=\left(\sum_{i=1}^5 J_i R_i(z) \tilde{J}_i\right)\left( 1 -
 K(z)\right)^{-1} \; .
\end{equation}
which is now to be transformed by the translation group $U(ib)$:
\begin{equation} \label{decoupling}
R(z;ib)=\left(\sum_{i=1}^5 J_i(ib) R_i(z;ib)
\tilde{J}_i(ib)\right)\left( 1 - K(z;ib)\right)^{-1}
\end{equation}
To prove that the eigenvalues of $H(F,ib)$ are at distance ${\cal
O}\left(\exp{(-1/F^{2(1-\varepsilon)})}\right)$ from those of
$H_2(F,ib)$, we have to show that the norm of $K(z;ib)$ becomes
smaller than $1$ as ${\rm dist}(\sigma(H_2(F)),z)$ becomes
Gaussian small. We will write $K(z;ib)$ as
\begin{equation} \label{normk}
K(z;ib)=\sum_{j=1}^5 K_j(z;ib) + M(z;ib)
\end{equation}
where
$$
K_j(z;ib)=[H_L,J_j(ib)]R_j(z;ib)\tilde{J}_j(ib)
$$
and
$$
M(z;ib)=\left(\sum_{j=1}^5 J_j(ib)\tilde{J}_j(ib)-1\right) -
A_1(ib) -A_5(ib) - A_2(ib)
$$

In Appendix A we estimate the norm of each term in the definition
of $K(z;ib)$ separately. Our strategy is the following. Each of $K_j(z;ib)$
can be viewed as an integral operator with the corresponding kernel of the
form $f({\bf x})G({\bf x},{\bf x}';z)h({\bf x}')$, where $G({\bf x},{\bf
  x}';z)$ is the Green function of $H_1$. Typically, the overlap of the
functions $f(x)$ and $h(x')$ decreases as $F\to 0$. Fact, which together
with the Gaussian decay of $G({\bf x},{\bf x}';z)$ at large distances, see
Appendix A, assures that the norm of each of $K_j(z;ib)$ will tend to zero
in the limit $F\to 0$. As for the operator $M(z;ib)$, we will see that for
small values of $F$ its norm can be made arbitrarily small by a proper choice
of the parameters of the decoupling functions.\\

The results of Appendix A yield the following estimate on
the norm of $K(z;ib)$
\begin{eqnarray}\label{knorm1}
\|K(z;ib)\| &\leq& {\cal C}\, F^{-{\cal C}}\,
\beta(z)^{-\sigma(\Re z)} \left(e^{-\frac{\beta(z)}{F^\tau}}
+e^{-B\frac{{\cal C}'(B,\Re z)}{ F^{2(1-\varepsilon)}}} \right)
\left
(1+\|R_2(z;ib)\| \right )\nonumber \\
&+&{\cal C}e^{-\frac{\tilde{{\cal
C}}}{F^{2(1-\varepsilon)}}}\left(\|R_1(z;ib)\|+\|R_2(z;ib)\|+1\right)
\end{eqnarray}
with ${\cal C}'(B,\Re z)=B c(\Re z)\to 0$ as $\Re z\to \infty$,
$\tilde{{\cal C}}$ depending on the decoupling scheme (in
particular we can set $\tilde{{\cal C}}=B\tilde{c}$),
$\beta(z)=\frac{\Im z+bF}{2F}$ and $\sigma(\Re z) \geq 1$
($\sigma(\Re z)\to \infty$ as $\Re z\to \infty$). We remark that
for $F<1$ we have $\beta(z)\leq \dist(\sigma(H_1(ib)),z)$.
Using the inequality
\begin{equation}\label{R_1} \|R_1(z;ib)\| \leq
\frac{1}{\dist(z,\overline{\Theta(H_1(ib))})} =
\frac{1}{\dist(z,\R-ibF)},
\end{equation}
where $\Theta(H_1(ib))$ is the numerical range of $H_1(ib)$, see
\cite[Prop. 19.7]{HSigal}, we can rewrite \eqref{knorm1} as in the
following Lemma:

\begin{lem}\label{knorm}
Let $F$ be small enough. Then for a given $z\in\C$ there exist
positive numbers ${\cal C}_1,\, {\cal C}_2$, $\sigma(\Re z)\geq1$
and ${\cal C}(B,\Re z)>0$, with ${\cal C}(B,\Re z)=Bc(\Re z)\to 0$
as $\Re z\to \infty$, such that
\begin{eqnarray}
\|K(z;ib)\| &\leq& {\cal C}_1\, F^{-{\cal C}_2}\,
\dist(\sigma(H_1(ib)),z)^{-\sigma(\Re z)}
\left(e^{-\frac{\dist(\sigma(H_1(ib)),z)}{F^\tau}}
+e^{-\frac{{\cal C}(B,\Re z)}{F^{2(1-\varepsilon)}}}
 \right ) \nonumber\\
&& \times  \left (1+\|R_2(z;ib)\| \right ).
\end{eqnarray}
\end{lem}

\section{Main result}\label{location2}

Armed with Lemma \ref{knorm} we are ready to prove an estimate on
the difference between the spectral projectors of $H(F,ib)$ and
$H_2(F,ib)$.

Let $\Gamma(e_\alpha)$ the path in the complex plane enclosing the
eigenvalue $e_\alpha\in \sigma(H(0))$ at finite distance to the
Landau levels (see Figure 3). More precisely
\begin{eqnarray}\label{newcircle}
\Gamma(e_\alpha) &:= & \Gamma_1(e_\alpha)\cup
\Gamma_2(e_\alpha)\cup\Gamma_3(e_\alpha)\cup\Gamma_4(e_\alpha)
\nonumber\\
\Gamma_1(e_\alpha) & := & \{\xi\in \C:\,
\Re\xi=e_{\alpha}-cF^{\varepsilon/2},\,
|\Im\xi|\leq \rho\} \nonumber\\
\Gamma_2(e_\alpha) & := & \{\xi\in \C:\,
\Re\xi=e_{\alpha}+cF^{\varepsilon/2},\,
|\Im\xi|\leq \rho\} \nonumber\\
\Gamma_3(e_\alpha) & := & \{\xi\in \C:\,
e_{\alpha}-cF^{\varepsilon/2}\, \leq \Re\xi\leq\,
e_{\alpha}+cF^{\varepsilon/2},\, \Im\xi = \rho\}
\nonumber\\
\Gamma_4(e_\alpha) & := & \{\xi\in \C:\,
e_{\alpha}-cF^{\varepsilon/2}\, \leq \Re\xi\leq\,
e_{\alpha}+cF^{\varepsilon/2},\, \Im\xi = -\rho\} \; .
\end{eqnarray}

\begin{figure}[!h]\label{gamma}
\begin{center}
\input{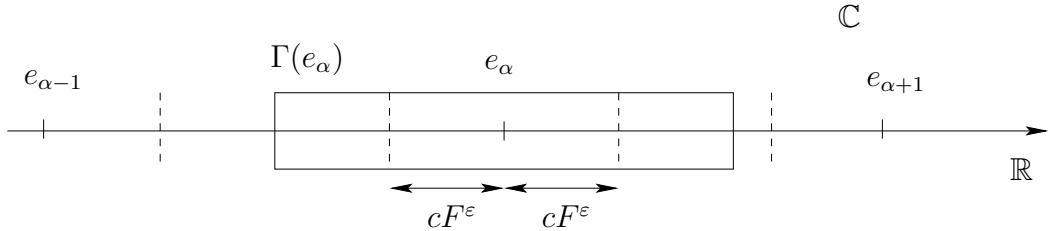}
\end{center}
\caption{\emph{The path $\Gamma(e_\alpha)$ in the complex plane.
The spectrum of $H_2(F,ib)$ is localised in the vicinity of
$e_\alpha$, represented by the dashed vertical lines.
$($Proposition \ref{prop21}$)$.}}
\end{figure}

For $F$ sufficiently small this construction can be made in such
a way that the spectrum of $H_2(F,ib)$ enclosed by
$\Gamma(e_\alpha)$ consists only of the eigenvalues
$\lambda_{\alpha,i}(F)\to e_\alpha$, where $i$ denote the
degeneracy index of the eigenvalue $e_\alpha$ ($1\leq i\leq
r_\alpha$), see Proposition \ref{prop21}. Moreover for $z\in
\Gamma(e_\alpha)$ holds by Lemma \ref{R_2}
\begin{equation}\label{r2dim}
\|R_2(z;ib)\|\leq {\cal C}\rho^{-s}\; .
\end{equation}
To control the inverse $(1-K(z,ib))^{-1}$ we need $\|K(z;ib)\|<1$ for $z\in
\Gamma(e_\alpha)$. In particular we want $\|K(z;ib)\|\to 0$ as
$F\to 0$. Looking at Lemma \ref{knorm}, together with
\eqref{r2dim} we see that the above requirement on the norm of
$K(z;ib)$ is satisfied at best taking
\begin{equation}\label{rho0}
\rho=e^{-\frac{\rho_0}{F^{2(1-\varepsilon)}}} \quad \text{with }
s\rho_0<{\cal C}(B,\Re z)
\end{equation}
We point out that the Gaussian smallness of $\rho$ is the optimal
choice to get the eigenprojectors convergence.
 \noindent From the decoupling formula \eqref{decoupling} we
have
\begin{eqnarray}\label{dec}
&& R(z;ib)-R_2(z;ib) = \Le(\sum_{i=1}^5
J_i(ib)R_i(z;ib)\tilde{J}_i(ib)\Ri)
\sum_{n=1}^\infty K(z;ib)^n -(1-J_2(ib))R_2(z;ib) \nonumber\\
&& - J_2(ib)R_2(z;ib)(1-\tilde{J}_2(ib)) + \sum_{i\in
\{1,3,4,5\}}J_i(ib) R_i(z;ib) \tilde{J}_i(ib) \; .
\end{eqnarray}
Because of $\sigma(H_i(ib))=\R-ibF$ (see \eqref{specH1nd}),
$R_i(z;ib)$, $i\not=2$, have no poles in $\Gamma(e_\alpha)$.
Moreover the only poles of $R_2(z;ib)$ are precisely
$\lambda_{\alpha,i}(F)$ ($1\leq i \leq r_\alpha$). Thus
integrating \eqref{dec} along the path $\Gamma(e_\alpha)\equiv
\Gamma$
\begin{eqnarray}\label{dpr}
P^{\Gamma}(ib)-P^{\Gamma}_2(ib) &=& \frac{1}{2\pi i}\oint_{\Gamma}
\Le(\sum_{i=1}^5 J_i(ib)R_i(z;ib)\tilde{J}_i(ib)\Ri)
\sum_{n=1}^\infty K(z;ib)^n \D z\nonumber\\
&-&J_2(ib)P^{\Gamma}_2(ib)(1-\tilde{J}_2(ib))-(1-J_2(ib))P^{\Gamma}_2(ib)
\;.
\end{eqnarray}
where $P^{\Gamma}_2(ib)$ is the spectral projector of $H_2(ib)$
onto $\overline{\Int \Gamma}$ and
$$
P^{\Gamma}(ib)= \frac{1}{2\pi i}\oint_{\Gamma} (z-H(ib))^{-1} \D z
$$
We estimate the norms of the three contributions on the r.h.s. of
\eqref{dpr}. If $\rho_0$ in the definition of $\Gamma(e_\alpha)$
satisfies a bit stronger condition than the bound in
\eqref{rho0}, the norm of the first term is smaller than
\begin{eqnarray}\label{a}
{\cal C}\Le(\sum_{i=1}^5\sup_{z\in \Gamma}\|R_i(z;ib)\|\Ri)
\frac{\sup_{z\in \Gamma} \|K(z;ib)\|}{1- \sup_{z\in \Gamma}
\|K(z;ib)\|} \leq g(F)  \to 0 \; \;\textrm{ as $F\to 0$}\; .
\end{eqnarray}
Indeed, for $i=2$, by \eqref{r2dim} and \eqref{rho0} there exists
a smooth function $g(F)$ such that
$$
\|R_2(z;ib)\|\|K(z;ib)\|\leq {\cal C}g(F)
$$
for each $z\in \Gamma(e_\alpha)$ and $\lim_{F\to 0}g(F)=0$
provided $2 s\rho_0 < {\cal C}(B,\Re z)$. For $i\not=2$
remembering that $b=b_0F^{\alpha}$, by \eqref{R_1} we have
$\sup_{z\in \Gamma}\|R_i(z;ib)\|\leq \frac{{\cal
C}}{F^{\alpha+1}}$, and the result follows.

To estimate the second term in \eqref{dpr} we write
\begin{eqnarray}
\|J_2(ib)P^{\Gamma}_2(ib)(1-\tilde{J}_2(ib))\| &\leq&
\|J_2(ib)\|_\infty \|P^{\Gamma}_2(ib)(1-\tilde{J}_2(ib))\| \nonumber\\
&\leq& \|[P^{\Gamma}_2(ib)-P^{\Gamma}_2(0)](1-\tilde{J}_2(ib))\|\nonumber\\
&+&\|[P^{\Gamma}_2(0)-P^{\Gamma}](1-\tilde{J}_2(ib))\|
+\|P^{\Gamma}(1-\tilde{J}_2(ib))\|\nonumber\\
&\leq&\left(\|P^{\Gamma}_2(ib)-P^{\Gamma}_2(0)\|+\|P^{\Gamma}_2(0)-
P^{\Gamma}\|\right)\|(1-\tilde{J}_2(ib))\|_\infty\nonumber\\
&+& \sum_{i=1}^{r_\alpha}|(1-\tilde{J}_2(ib),\phi_{0}^i)|
\end{eqnarray}
where $P^{\Gamma}$ is the spectral projector of $H(0)$ onto the
eigenfunctions $\phi_0^i$ ($i=1,\ldots,r_\alpha$) corresponding to
the eigenvalue $e_\alpha$. In order to control the term
$\|P^{\Gamma}_2(ib)-P^{\Gamma}_2(0)\|$ we define a circle
$\tilde{\Gamma}\equiv\{\xi\in \C:\, |\xi-e_\alpha|=\Gamma_0
F^\varepsilon \}$. Then for $F$ small enough holds
\begin{eqnarray}\label{deltap}
\|P^{\Gamma}_2(ib)-P^{\Gamma}_2(0)\| &\leq&
(2\pi)^{-1}\oint_{\tilde{\Gamma}}\|R_2(\xi;ib)A(ib)
R_2(\xi;0)\|\, |\D \xi|\nonumber \\
&\leq& {\cal C}\, F^{\alpha-1}
\end{eqnarray}
where $A(ib)$ is defined in \eqref{Aib} and the second inequality
follows form \eqref{normr2}. By Proposition \ref{prop21}
$\|P^{\Gamma}_2(0)-P^{\Gamma}\|\to 0$ as $F\to 0$. Thus for $F\to
0$ the two terms above are infinitesimally small. The last term
can be easily estimated using the result of \cite[Thm. 4.2]{CN},
which says that for any at least gaussian decaying potential one has the
estimate
$$
|\phi({\bf x})| \leq {\cal C} e^{-\mu|{\bf x}|^2},
$$
where $\phi$ is associated to a discrete eigenvalue of $H(0)$.
Using this result and a bound on $|1-\tilde{J}_2(ib)|$ similar to
that of \eqref{j0b} we get
\begin{equation}\label{bbb}
\|J_2(ib)P^{\Gamma}_2(ib)(1-\tilde{J}_2(ib))\| \to 0 \qquad
\textrm{as $F\to 0$}
\end{equation}
For the third term in \eqref{dpr} we
obtain the same estimate, since $\|A^*\|=\|A\|$. 
In conclusion we arrive at

\begin{prop}\label{dimPH}
Let $\Gamma(e_\alpha)$ be as in \eqref{newcircle}, then
$$
\|P^{\Gamma}(ib)-P^{\Gamma}_2(ib)\| \to 0, \qquad F\to 0
$$
In other words, $\dim \Ran P^{\Gamma}(ib)=\dim \Ran
P^\Gamma_2(ib)$ for $F$ sufficiently small.
\end{prop}

Propositions \ref{dimPH} and \ref{prop21} yield

\begin{thm} \label{convergence}
Assume $V$ satisfies $(a),\, (b)$ and let $e_\alpha$ be the
eigenvalue of $H(0)$ of multiplicity $r_\alpha$. Then near
$e_\alpha$ there are eigenvalues $E_{\alpha,i}$ of $H(F,ib),
(1\leq i \leq r_\alpha)$, repeated according to their
multiplicity, and
$$
E_{\alpha,i}\ra e_\alpha \quad {\rm as}\quad F\ra 0.
$$
\end{thm}

\noindent Now we can formulate our main result.

\begin{thm} \label{main}
Assume $V$ satisfies $(a)$ and $(b)$. Let $e_\alpha$ and
$E_{\alpha,i}$ be the eigenvalues defined in Theorem
\ref{convergence}. Then there exist some positive constants ${\cal
C}$ and $R_\alpha(B)$, such that for $F$ small enough the
following inequality holds true
$$
|\Im E_{\alpha,i}|\leq {\cal C}\,
e^{-\frac{R_\alpha(B)}{F^{2(1-\varepsilon)}}}, \quad \varepsilon>0,
$$
where $\varepsilon$ can be made arbitrarily small and
$R_\alpha(B)=B\tilde{R}_\alpha$.
\end{thm}

\begin{proof}
Consider the path $\Gamma(e_\alpha)$ defined through
(\ref{newcircle}), with $\rho_0=R_\alpha(B)$. We have proved in
Proposition \ref{dimPH} that if
\begin{equation}\label{depRa}
2s\, R_\alpha(B)<{\cal C}(B,e_\alpha),
\end{equation}
with ${\cal C}(B,e_\alpha)$ defined in Lemma \ref{knorm}, then
$\dim \Ran P^\Gamma(ib)=\dim \Ran P_2^\Gamma(ib)$ and the only
eigenvalues of $H(F,ib)$ in $\overline{\Int \Gamma}$ are the
eigenvalues $E_{\alpha,i}$. By construction their imaginary parts
satisfy the announced upper bound. The linear dependence on $B$
follows from the linear dependence of ${\cal
C}(B,e_\alpha)$ on $B$.
\end{proof}

\begin{rem}
The behaviour of $\tilde{R}_\alpha$ w.r.t. $\alpha$ is not
uniform. Indeed $\tilde{R}_\alpha\to 0$ as $e_\alpha\to \infty$,
because ${\cal C}(B,\Re z)\to 0$ as $\Re z \to \infty$.
\end{rem}

As already mentioned at the end of Section \ref{modelres2} the resonance
widths are given by the imaginary parts of the eigenvalues of $H(F,ib)$, and the
lifetime by the inverse of the resonance width. Since
$\varepsilon$ is arbitrarily small, we thus get a lower bound on the
life-times:

\begin{cor}
The life-times of the resonant states satisfy:
$$
\tau_\alpha = \tfrac{1}{2}\sup_{\varepsilon>0} |\Im
E_{\alpha,i}|^{-1} \geq 1/{\cal C}\,
\exp\left(\frac{B\tilde{R}_\alpha}{F^2}\right)\; .
$$
\end{cor}

\subsubsection*{Conclusion}

Theorem \ref{main} gives a partial generalisation of the result
obtained in \cite{GM}. As expected, the fact that the lower bound
on the resonance life-times is Gaussian in $F^{-1}$ and not
exponential is due to the presence of the magnetic field. However,
further comparison with the purely electric Stark effect shows
much larger restriction on the class of admissible potentials, in
particular the condition on the Gaussian decay of $V(x,y)$. Let us
now briefly discuss the issue of Gaussian versus exponential
behaviour. As follows from the analysis of the Stark resonances,
\cite{oppen} \cite{HSimon} \cite{sig}, the exponential law for the
resonant states is in that case directly connected with the
exponential decay of the eigenfunctions of a ``free'' Hamiltonian,
i.e. without electric field. If we suppose that the same
connection exists also in the magnetic case, then our result
should hold whenever the eigenfunctions of $H(0)=H_L+V$,
associated with the discrete spectrum, fall off as a Gaussian.
Sufficient condition for the latter is the Gaussian decay of
$V(x,y)$, see \cite{CN}, which is compatible with our assumption
$(b)$. Up to now, the optimal condition is known only for the
ground state, in which case a sort of exponential decay of
$V(x,y)$ is shown to be sufficient and necessary for Gaussian
behaviour of the corresponding eigenfunctions at infinity,
\cite{Erdos}. \par Such a restriction is in contrast with the non
magnetic Schr\"odinger operator, whose eigenfunctions decrease
exponentially in the classically forbidden region independently on
the rate at which $V(x,y)$ tends to zero at infinity. This might
indicate a principal difference between the behaviour of resonant
states in the presence respectively absence of magnetic field.

\appendix

\section{Estimate of $\|K(z;ib)\|$}

Here we estimate the norm of each term in the definition of
$K(z;ib)$ separately. Since the calculations are often analogous,
we skip the details in many places.

\subsubsection*{Norm of $M(z;ib)$}

\noindent  Terms $\|A_1(ib)\|$ and $\|A_5(ib)\|$:
\begin{eqnarray}
\|A_1(ib)\| &\leq& \|V(ib)J_1(ib)\|_\infty
\|R_1(z;ib)\|\|\tilde{J}_1(ib)\| \nonumber\\
& \leq &{\cal C} \|V(ib)J_1(ib)\|_\infty \|R_1(z;ib)\|
\end{eqnarray}
and for $F$ sufficiently small
\begin{eqnarray*}
\|V(ib)J_1(ib)\|_\infty &=&
\sup_{(x,y)}|V(x+ib,y)||J_-(x+ib)||J_c(y)| \\
&\leq&
\sup_{x}|V(x+ib,\hat{y})|\frac{e^{2\gamma(x-x_2)}}{\left(e^{4\gamma(x-
x_2)}+1\right)^{1/2}}
\end{eqnarray*}
We estimate this term as $\max \{a,b,c\}$ where $a$, $b$, $c$ are
\begin{eqnarray*}
a&=& \sup_{|x|<a_0} |V(x+ib,\hat{y})| e^{2\gamma(x-x_2)} \leq
V_0e^{2\gamma(a_0-x_2)} 
\leq V_0e^{\frac{2\gamma_0
a_0}{F^{1-\varepsilon}}}e^{-\frac{2\gamma_0
C_2}{F^{2(1-\varepsilon)}}} \\
b&=& \sup_{a_0\leq |x|\leq a_0+\delta} V_0e^{-\nu x^2}
e^{2\gamma(x-x_2)} \leq
V_0e^{-\nu a_0^2} e^{2\gamma(a_0+\delta-x_2)}
\leq V_0e^{\frac{2\gamma_0 a_0}{F^{1-\varepsilon}}}e^{-
\frac{2\gamma_0(C_2-\delta_0)}{F^{2(1-\varepsilon)}}}\\
c&=& \sup_{|x|>a_0+\delta}V_0e^{-\nu x^2} \leq
V_0e^{-\frac{{\delta_0}^2}{F^{2(1-\varepsilon)}}}
\end{eqnarray*}
and $\delta=\delta_0F^{-(1-\varepsilon)}<x_2$. This leads to
$$
\|A_1(ib)\|\leq {\cal C} e^{-\frac{{\cal
C}}{F^{2(1-\varepsilon)}}} \|R_1(z;ib)\|
$$
In the same way we prove the estimate for $\|A_5(ib)\|$.\\

\noindent  Term $\|A_2(ib)\|$:
\begin{eqnarray}
\|A_2(ib)\| &\leq& F\|(x+ib)(1-h_F(x+ib))J_2(ib)\|_\infty
\|R_2(z;ib)\|\|\tilde{J}_2(ib)\| \nonumber\\
& \leq &{\cal C}F \|(x+ib)(1-h_F(x+ib))J_0(x+ib)\|_\infty
\|R_2(z;ib)\|
\end{eqnarray}
We can easily found the following bounds
\begin{equation}\label{j0b}
|J_0(x+ib)| \leq  \frac{1}{\cos(2\gamma b)} \begin{cases}
e^{2\gamma(x+x_1)} &\quad
\textrm{ if } x<0\\
e^{-2\gamma(x-x_1)} &\quad \textrm{ if } x>0
\end{cases}
\end{equation}
and
\begin{equation}\label{hFb}
|1-h_F(x+ib)| \leq \left(e^{-4\gamma(x-\bar{x})}+1\right)^{-1/2}+
\left(e^{4\gamma(x+\bar{x})}+1\right)^{-1/2} \equiv h_1+h_2
\end{equation}
For $x>\frac{\bar{x}+x_1}{2}>0$
\begin{eqnarray*}
|h_1|^2|J_0(x+ib)|^2 &\leq&{\cal
C}\frac{e^{-4\gamma(x-x_1)}}{e^{-4\gamma(x-\bar{x})}+1} \leq{\cal
C}\frac{e^{-4\gamma(x-\frac{\bar{x}+x_1}{2})}}{e^{-2\gamma(x_1-\bar{x})}}
\end{eqnarray*}
the last inequality follows after multiplication by
$(e^{2\gamma(\bar{x}-x_1)})/(e^{2\gamma(\bar{x}-x_1)})$. Now,
$y=x-(\bar{x}+x_1)/2$, yields
\begin{eqnarray}
\sup_{x>\frac{\bar{x}+x_1}{2}} F|x||h_1J_0(x+ib)| &\leq& {\cal C}
F \sup_y \left(|y|+|\bar{x}+x_1|/2\right)
e^{-\gamma(\bar{x}-x_1)}e^{-2\gamma|y|}\nonumber \\
&\leq& {\cal C}(F+F^{\varepsilon}) e^{-\frac{{\cal
C}}{F^{2(1-\varepsilon)}}} \label{qay}
\end{eqnarray}
For $x<-\frac{\bar{x}+x_1}{2}<0$ we get in the same way the upper
bound \eqref{qay}. Finally, for $|x|\leq \frac{\bar{x}+x_1}{2}$
obviously $\sup _x |x| =\frac{\bar{x}+x_1}{2}$ and
$$
|h_1J_0(x+ib)| \leq e^{-2\gamma (\bar{x}-x_1)}
$$
which gives a similar estimate as \eqref{qay}. \\
A similar argument holds for $|h_2J_0(x+ib)|$ that leads to
\begin{equation}\label{xxx}
\|A_2(ib)\| \leq {\cal C} e^{-\frac{{\cal
C}}{F^{2(1-\varepsilon)}}}  \|R_2(z;ib)\|
\end{equation}

\noindent Term $\|\sum_{j=1}^5
J_j(ib)\tilde{J}_j(ib)-1\|$: \\
First we remark that we can write
$1=\tilde{J}_c(y)+(1-\tilde{J}_c(y))$ and that $\sum_{i=3}^4
J_i(ib)\tilde{J}_i(ib) -(1-\tilde{J}_c)=0$, thus it remains to
estimate $\sum_{i\in\{1,2,5\}}
J_i(ib)\tilde{J}_i(ib)-\tilde{J}_c$. We have
\begin{eqnarray*}
\sum_{i\in\{1,2,5\}} J_i(ib)\tilde{J}_i(ib)-\tilde{J}_c &=&
\left[J_-(x+ib)\tilde{J}_-(x+ib)+J_0(x+ib)\tilde{J}_0(x+ib)\right.
\\
&+&\left. J_+(x+ib)\tilde{J}_+(x+ib)-1\right]\tilde{J}_c(y)
:={\cal X}(ib)\tilde{J}_c(y)
\end{eqnarray*}
Now $\|\tilde{J}_c(y)\|_\infty=1$, and it remain to estimate
\begin{equation}\label{yyxc}
\|{\cal X}(ib)\|_\infty = \left\|\sum_{\alpha\in\{\pm,0\}}
J_\alpha(x)\tilde{J}_\alpha(x)-1 \right\|_\infty
\end{equation}
This can be done by developing explicitly the functions in term of
the exponentials and write the sum as fraction (denote by $\cal K$
the denominator). After a tedious straightforward computation we
find out that each term in the sum
$$
\sum_{\alpha\in\{\pm,0\}} J_\alpha(x+ib)\tilde{J}_\alpha(x+ib)-1
$$
can be bounded from above uniformly w.r.t. $x$ by ${\cal C}
e^{-{\cal C}F^{-(2-\varepsilon)}}$. For example
$$
\left|\frac{e^{-2\gamma(2x+x_0+x_2)}}{{\cal K}}\right| \leq
\frac{e^{-2\gamma(2x+x_0+x_2)}}{\cos(4\gamma b)e^{4\gamma x}} =
\frac{e^{-2\gamma(x_0+x_2)}}{\cos(4\gamma b)} \leq {\cal
C}e^{-\frac{{\cal C}}{F^{2(1-\varepsilon)}}}
$$
for $F\to 0$ due to \eqref{bsmall} and similarly in other cases.
Therefore
$$
\left\|\sum_{i=1}^5 J_i(ib)\tilde{J}_i(ib) -1\right\|_\infty \leq
{\cal C}e^{-\frac{{\cal C}}{F^{2(1-\varepsilon)}}}
$$

\noindent
 Finally,
$$
\|M(z;ib)\| \leq  {\cal C}e^{-\frac{{\cal
C}}{F^{2(1-\varepsilon)}}}\left(\|R_1(z;ib)\|+\|R_2(z;ib)\|+1\right)
$$

\subsubsection*{Norm of $K_3(z;ib)$ and $K_4(z;ib)$}

To control the operator norm we will use alternatively the
Hilbert-Schmidt norm and the following inequality for the norm of
an integral operator which can be found in \cite[p. 144]{Kato}
\begin{equation} \label{maxnorm}
\|A\| \leq \max \left\{\sup_{{\bf x}} \int |A({\bf x},{\bf x}')|
\D {\bf x} ; \sup_{{\bf x}'} \int |A({\bf x},{\bf x}')| \D {\bf x}
 \right\}
\end{equation}
Each integration that we need to evaluate is split in two parts
according to $|x-x'|\geq 1$ and $|x-x'|<1$: \par \noindent Let
$\varphi$ such that $\|\varphi\|=1$, and $A$ an operator with
integral kernel $A({\bf x},{\bf x}')$, then
\begin{eqnarray}
\|A\varphi\|^2 &=&\int_{\R^2} \left| \int_{\R^2} A({\bf
x},{\bf x}')\varphi({\bf x}')\D {\bf x}' \right|^2  \D {\bf x} \\
&\leq& 2\int_{\R^2} \left| \int_{\substack{\R^2:}{|x-x'|\geq 1}}
A({\bf x},{\bf x}')\varphi({\bf x}')\D {\bf x}' \right|^2\D {\bf
x} \nonumber \\
&& + 2\int_{\R^2} \left| \int_{\substack{\R^2:}{|x-x'|< 1}} A({\bf
x},{\bf x}')\varphi({\bf x}')\D {\bf x}'\right|^2\D {\bf x} =:
2(a+b)\, .
\end{eqnarray}
We now treat the two terms separately. By the Schwartz inequality
we have
$$
a\leq \int_{\R^2} \int_{\substack{\R^2:}{|x-x'|\geq 1}}|A({\bf
x},{\bf x}')|^2\D {\bf x}' \D {\bf x}\|\varphi\|^2 \leq
\|A\|_{HS}^2\|\varphi\|^2
$$
For $b$ we proceed as follows, let
$$
\psi({\bf x})\equiv \int_{\substack{\R^2:}{|x-x'|< 1}} A({\bf
x},{\bf x}')\varphi({\bf x}')\D {\bf x}'
$$
and
$$
A({\bf x})= \int_{\substack{\R^2:}{|x-x'|< 1}} |A({\bf x},{\bf
x}')|\D {\bf x}' \qquad A'({\bf x}')=
\int_{\substack{\R^2:}{|x-x'|< 1}} |A({\bf x},{\bf x}')|\D {\bf x}
$$ we first remark that $\int_{\substack{\R^2:}{|x-x'|< 1}}
|A({\bf x},{\bf x}')|/A({\bf x})\D {\bf x}'=1$, this implies by
convexity, that
$$
\left(\frac{|\psi({\bf x})|}{A({\bf x})}\right)^2 \leq
\int_{\substack{\R^2:}{|x-x'|< 1}} \frac{|A({\bf x},{\bf
x}')|}{A({\bf x})}|\varphi({\bf x}')|^2\D {\bf x}'
$$
and thus
\begin{eqnarray}
b=\int_{\R^2}  |\psi({\bf x})|^2\D {\bf x} &\leq& \sup_{{\bf x}}
A({\bf x}) \int_{\R^2} \int_{\substack{\R^2:}{|x-x'|< 1}}
|A({\bf x},{\bf x}')||\varphi({\bf x}')|^2\D {\bf x}'\D {\bf x} \nonumber\\
&=&\sup_{{\bf x}} A({\bf x}) \int_{\R^2}
\int_{\substack{\R^2:}{|x-x'|< 1}}
|A({\bf x},{\bf x}')||\varphi({\bf x}')|^2\D {\bf x} \D {\bf x}'\nonumber\\
&\leq& \sup_{{\bf x}} A({\bf x})\sup_{{\bf x}'} A'({\bf x}')
\|\varphi\|^2\nonumber\\
&\leq& \max \left\{\sup_{{\bf x}} A({\bf x}),\sup_{{\bf x}'}
A'({\bf x}') \right\}^2 \|\varphi\|^2 \label{maxnormbis} \eea
Therefore, for $|x-x'|\geq 1$ we can use a Hilbert-Schmidt-like
norm, while for $|x-x'|<1$ we can use a \eqref{maxnorm} norm. We
will need results on the behaviour of the Green function $G_1({\bf
x},{\bf x}';z)$ of $H_1(ib)$. We expect that at points ${\bf
x},{\bf x}'$ with $|x-x'|$ large the Green function decay in the
$x-$direction as a Gaussian due to the magnetic field, while in
the $y-$direction (the drift direction of the classical particle)
we expect only exponential decay. On the other we also expect
integrable singularity at the origin. These properties are
contained in the following two lemmas which are obtained in
\cite{FK1}.

\begin{lem}\label{greenasympt}
Let $|x-x'|\geq 1$ and let $F$ be small enough. Then there exist
some strictly positive constants $G_0, \, \omega(z)$ and
$\sigma(z)\geq 1$ such that $$|\pd^n_{x,y} G_1({\bf x},{\bf
x}';z)|\leq G_0\, \beta(z)^{-\sigma(z)}\, e^{-\beta(z)
  |y'-y|}\, e^{-\omega(z)(x'-x)^2},
$$where $n=0,1$ and $\beta(z)=\frac{\Im z+bF}{2F}$.
\end{lem}

\begin{lem}\label{greenasympt1}
For $F$ small enough there exists some strictly positive constants
$G'_0$ and $\sigma(z)$, such that the following inequality holds
true
\begin{equation}
\int_{\R}\int_{|x'-x|< 1} |\pd_{x,y}^n G_1({\bf x},{\bf x}';z)|
e^{\frac{\beta(z)}{2}|y-y'|} \,{\rm d}{x'} {\rm d}{y'} \leq G'_0\,
\beta(z)^{-\sigma(z)},
\end{equation}
where $n=0,1$ and $\beta(z)=\frac{\Im z+bF}{2F}$.
\end{lem}

Since the integrands are positive functions, for $|x-x'|\geq 1$ we
first substitute the integral kernels by their upper bounds and
then integrate without any restriction.

\begin{rem}\label{remunif}
In the Lemmas above the coefficient $\omega(z)$ depends only in
$\Re z$ and decreases as $\Re z$ increases.  Moreover, $\omega(z)$
is linear in $B$: $\omega(z)\sim B$. $\sigma(z)\geq 1$, and also
depends only on $\Re z$ and diverges for $\Re z\to \infty$. For
the sake of brevity we do not write $z$ in the arguments of
$\sigma$ and $\omega$.
\end{rem}

We now evaluate the norm of $K_3(z;ib)$. The terms in the
commutator are
$$
[p_y^2,J_3(ib)]R_3(z;ib)\tilde{J}_3(ib)=-2\pd_x J_3(ib)\pd_x
R_3(z;ib)\tilde{J}_3(ib) -\pd_x^2 J_3(ib)R_3(z;ib)\tilde{J}_3(ib)
$$

We use again inequality \eqref{maxnorm}. Due to the upper bound on
the Green function and its derivatives when $|x-x'|\geq 1$ the
integration can be separated in two parts, which for $F$ small
enough gives us (for $n=1,2$)
\begin{eqnarray*}
& &\sup_{{\bf x}} \int \D {\bf x}'
|\pd_y^nJ_3(x+ib,y)||\pd_y^{2-n}G_3
({\bf x},{\bf x}';z)||\tilde{J}_3(x'+ib,y')| \\
&\leq& {\cal C} \sup_y \int dy' |\pd_y^nJ_>
(y)|\beta(z)^{-\sigma}e^{-\beta(z)|y-y'|}|\tilde{J}_>(y')| \\
&\leq&  {\cal C} \beta(z)^{-\sigma}\sup_{y\in \supp \pd_y^nJ_>} \;
\sup_{y'\in \supp \tilde{J}_>} e^{-\frac{\beta(z)}{2}|y-y'|} =
{\cal C}\beta(z)^{-\sigma}e^{-\frac{\beta(z)}{2F^\tau}}
\end{eqnarray*}
and similarly for the second term. We now consider the situation
$|x-x'|<1$, let be the set $D=\{x'\in \R:|x-x'|<1\}\times \R$
\begin{eqnarray*}
& &\sup_{{\bf x}} \int_{D} \D {\bf x}'
|\pd_y^nJ_3(x+ib,y)||\pd_y^{2-n}G_3
({\bf x},{\bf x}';z)||\tilde{J}_3(x'+ib,y')| \\
&\leq& \sup_{{\bf x}}\int_{D} \D {\bf x}'
|\pd_y^nJ_3(x+ib,y)|e^{-\frac{\beta(z)}{2}|y-y'|}
|\tilde{J}_3(x'+ib,y')| |\pd_y^{2-n}G_3 ({\bf x},{\bf
x}';z)|e^{\frac{\beta(z)}{2}|y-y'|}\\
&\leq&  \sup_{y\in \supp \pd_y^nJ_>} \; \sup_{y'\in \supp
\tilde{J}_>} e^{-\frac{\beta(z)}{2}|y-y'|} \sup_{{\bf x}}
 \int_{D} \D {\bf x}'|\pd_y^{2-n}G_3 ({\bf x},{\bf
x}';z)|e^{\frac{\beta(z)}{2}|y-y'|} \\
&\leq& {\cal C}\beta(z)^{-\sigma}e^{-\frac{\beta(z)}{2F^\tau}}
\end{eqnarray*}

Thus we can conclude that
$$
\|K_3(z;ib)\| \leq {\cal C}
\beta(z)^{-\sigma}e^{-\frac{\beta(z)}{2F^\tau}}
$$

\noindent In the same way we prove the estimate for
$\|K_4(z;ib)\|$.

\subsubsection*{Norm of $K_1(z;ib)$ and $K_5(z;ib)$}

Here below when we write $\|\cdot \|_{HS}$ for $|x-x'|\geq 1$ it
is understood that part of the Hilbert-Schmidt, which corresponds
to the integration over $\R^2$ with the restriction $|x-x'|\geq
1$. For the integral kernel of $R_1(z;ib)$ and $\pd_{x,y}\,
R_1(z;ib)$
we then use the upper bounds of Lemma \ref{greenasympt}. \\

The first term in the commutator $[H_L,J_1(ib)]$ gives
\begin{equation} \label{commut}
[p_x^2,J_1(ib)]R_1(z;ib)\tilde{J}_1(ib) = -2\pd_x J_1(ib)\pd_x
R_1(z;ib)\tilde{J}_1(ib) -\pd_x^2 J_1(ib)R_1(z;ib)\tilde{J}_1(ib)
\end{equation}
In the case $|x-x'|\geq 1$ we estimate the ``restricted''
Hilbert-Schmidt norms term by term.
\begin{eqnarray*}
& & \|\pd_x J_1(ib) \pd_x R_1(z;ib)\tilde{J}_1(ib)\|^2_{HS}=  \\
& = & \int_{\R^4}|J'_-(x+ib)J_c(y)|^2|\pd_x G_1({\bf x},{\bf
x}';z)|^2|\tilde{J}_-(x'+ib)\tilde{J}_c(y')|^2\D {\bf x}\D {\bf
x}'
\end{eqnarray*}
As before due to the properties of the Green function for
$|x-x'|\geq 1$ the integration can be separated it two parts. One
can easily check that the integral with respect to $y,y'$ gives
the factor
$$
{\cal C}\, F^{-2\tau}
$$
The second part is bounded above by
$$
\beta(z)^{-\sigma}\int_{\R} |J'_-(x+ib)|^2 f(x,x_0)\D x
$$
where
$$
f(x,x_0):= \int_{\R} e^{-\omega(x-x')^2}
\frac{1}{1+e^{-4\gamma(x'-x_0)}}\, \D x'
$$
Here we have used the fact that for $F$ sufficiently small (see
(\ref{bsmall}))
\begin{eqnarray}\label{diff0}
|\tilde{J}_-(x'+ib)|^2 & = & \left
(1+e^{-4\gamma(x'-x_0)}+2\cos(2\gamma
b)e^{-2\gamma(x'-x_0)}\right )^{-1}\nonumber \\
& \leq & \frac{1}{1+e^{-4\gamma(x'-x_0)}}
\end{eqnarray}
In the similar way we find out that
\begin{equation}\label{diff1}
|J'_-(x+ib)|^2\leq {\cal C}\, F^{-2} e^{-4\gamma|x-x_2|}
\end{equation}
so that it suffices to look for an upper bound on the functional
\begin{eqnarray} \label{3int}
& & \int_{\R} e^{-4\gamma|x-x_2|}f(x,x_0)\D x =
\int_{-\infty}^{x_2 -
\delta} e^{-4\gamma|x-x_2|}f(x,x_0)\D x \nonumber \\
& & + \int_{x_2 +\delta}^{\infty}  e^{-4\gamma|x-x_2|}f(x,x_0)\D x
+ \int_{x_2 -\delta}^{x_2 +\delta} e^{-4\gamma|x-x_2|}
f(x,x_0)\D x  \nonumber \\
& = & I_1+I_2+I_3
\end{eqnarray}
where $\delta=\delta_0 F^{-1(1-\varepsilon)}$ such that $(x_2
+\delta) < x_0$. As $f(x,x_0)$ is by definition strictly positive
and bounded, the first two integrals on the r.h.s. of (\ref{3int})
can be easily estimated as follows
\begin{eqnarray*}
I_1+I_2 & \leq  & e^{-2\gamma\delta} \|f\|_{\infty}\left
[\int_{-\infty}^{x_2 -\delta} e^{2\gamma(x-x_2)}\D x +
\int_{x_2 +\delta}^{\infty}  e^{-2\gamma(x-x_2)}\D x\right ] \\
& \leq & \gamma^{-1}\sqrt{\frac{\pi}{\omega}}e^{-2\gamma\delta}
\end{eqnarray*}
In order to control $I_3$ we have to look at the function
$f(x,x_0)$ in more detail. First we note that
\begin{eqnarray}\label{division}
 f(x,x_0) & = & \int_{\R} e^{-\omega(x-x_0-t)^2}\, \frac{\D t}{1+e^{-
4\gamma t}} \nonumber \\
& \leq & \int_0^{\infty} e^{-\omega(x-x_0-t)^2}\D
t+\int_{-\infty}^0 e^{-\omega(x-x_0-t)^2+4\gamma t}\D t
\end{eqnarray}
>From \cite[p. 1064]{GR} (see also \eqref{integral})  we then get
the bound on $f(x,x_0)$ in the form
\begin{eqnarray*}
f(x,x_0)& \leq & \sqrt{\frac{1}{2\omega}}\,
e^{-\omega(x-x_0)^2}\Bigg [e^{\frac{\omega(x-x_0)^2}{2}}
 D_{-1}(\sqrt{2\omega}\, (x_0-x))  \\
& + & e^{\frac{(2\omega(x-x_0)+4\gamma)^2}{8\omega}}
D_{-1}\left(\frac{2\omega(x-x_0)+4\gamma}{\sqrt{2\omega}}\right
)\Bigg ]
\end{eqnarray*}
where $D_{-1}(\cdot)$ denotes the parabolic cylinder function.
Using its asymptotic expansion \cite[p. 1065]{GR}
$$
\begin{array}{lclc}
D_{-1}(z) & = & e^{-z^2/4}z^{-1}(1-\mathcal{O}(z^{-2})), &
z\ra\infty
\\
D_{-1}(z) & = & e^{z^2/4}(1+\mathcal{O}(z^{-2})), & z\ra-\infty
\end{array}
$$
it is not difficult to verify that
$$
f(x,x_0)\leq {\cal C}e^{-{\cal C}\, F^{-2(1-\varepsilon)}},\quad
F\ra 0
$$
uniformly for any $x\in [x_2 -\delta,x_2 +\delta]$. Now we employ
the mean value theorem of the integral calculus which tells us
that there exists some $\tilde{x}\in [x_2 -\delta,x_2 +\delta]$
for which
$$
I_3=f(\tilde{x})\int_{x_2 -\delta}^{x_2
+\delta}e^{-4\gamma|x-x_2|}\, \D x = \frac{1}{2\gamma}\left
(1-e^{-4\gamma\delta}\right )f(\tilde{x})
$$
Let us remark that the second term of the commutator
(\ref{commut}) can be bounded in the same way, since
\begin{equation}\label{diff2}
|J''_-(x+ib)|^2\leq {\cal C}\, F^{-4}\, e^{-4\gamma|x-x_2|},\quad
F\ra 0
\end{equation}
Moreover, due to the decoupling with respect to $y-$axis, the
above procedure can be applied also to the second term in the
commutator $[H_L,J_1(ib)]$, namely
$$
[2Byp_x,J_1(ib)]R_1(z;ib)\tilde{J}_1(ib) = -2By\pd_x J_1(ib)
R_1(z;ib)\tilde{J}_1(ib)
$$
This allows us to find some $c_1(V,B)>0$ such that the following
holds true for $|x-x'|\geq 1$:
\begin{equation}\label{poi}
\left\|[(p_x+By)^2,J_1(ib)]R_1(z;ib)\tilde{J}_1(ib)\right
\|^2_{HS} \leq {\cal C}\, \beta(z)^{-\sigma}F^{- {\cal C}}\,
e^{-c_1(B)F^{-2(1-\varepsilon)}}
\end{equation}
where the constant $c_1(B)$ is proportional to $B$ (since
the factor $\omega$ is linear in $B$). \\
When $|x-x'|<1$ we use (\ref{maxnormbis}). As in the case
$|x-x'|\geq 1$ all the term in the commutator $[H_L,J_1(ib)]$
involving $x-$derivatives are treated in the same way. For example
for $\partial_xJ_1(ib)\partial_x R_1(z;ib)\tilde{J}_1(ib)$ we have
\begin{eqnarray} & &\sup_{{\bf x}} \int_{\R} \D y' \int_{x':|x-x'|<1} \D x'
|J_-'(x+ib)J_c(y)||\partial_xG_1({\bf x},{\bf
x}';z)||\tilde{J}_-(x'+ib)\tilde{J}_c(y')|\nonumber \\
&\leq&\sup_{{\bf x}}
\sup_{x':|x-x'|<1}|J_-'(x+ib)\tilde{J}_-(x'+ib)| \int_{\R} \D y'
\int_{x':|x-x'|<1} \D x'
|\partial_xG_1({\bf x},{\bf x}';z)|\nonumber \\
&\leq& {\cal C}\beta(z)^{-\sigma}
\sup_{x',x:|x-x'|<1}|J_-'(x+ib)\tilde{J}_-(x'+ib)| \label{petit1}
\eea and similarly for ${\bf x}$ and ${\bf x}'$ interchanged. Now,
using \eqref{diff0} and \eqref{diff1}, we get
$$
\sup_{x',x:|x-x'|<1}|J_-'(x+ib)\tilde{J}_-(x'+ib)| \leq {\cal
C}F^{-1}\sup_{x}
\frac{e^{-4\gamma|x-x_2|}}{1+e^{-4\gamma(x-x_0)}}\leq {\cal
C}F^{-1}e^{-{\cal C}F^{-2(1-\varepsilon)}} $$ This with
\eqref{poi} leads to
$$
\left\|[(p_x+By)^2,J_1(ib)]R_1(z;ib)\tilde{J}_1(ib)\right \|^2
\leq {\cal C}\, \beta(z)^{-\sigma}F^{- {\cal C}}\,
e^{-c_2(B)F^{-2(1-\varepsilon)}}
$$
for $c_2(B)>0$.\\
To control the operator norm of the last term in the commutator
$[H_L,J_1(ib)]$, namely
$$
[p_y^2,J_1(ib)]R_1(z;ib)\tilde{J}_1(ib)
$$ we use again the inequality (\ref{maxnorm}). When
$|x-x'|\geq 1$, since both $f(x,x_0)$ and $f(x,x_2)$ are bounded
as well as $J_-(x+ib),\, \tilde{J}_-(x+ib)$, it suffices to
estimate these parts in (\ref{maxnorm}) which correspond to the
integration w.r.t. $y,y'$:
\begin{eqnarray}
& & \sup_y
|J'_c(y)|\int_{\R}e^{-\beta(z)|y-y'|}|\tilde{J}_c(y')|\, \D y'\leq
\sup_y|J'_c(y)|
\int_{-y_0}^{y_0}e^{-\beta(z)|y-y'|}\, \D y' \nonumber \\
& & \leq 2y_0\|J'_c\|_{\infty}\, e^{-\beta(z)F^{-\tau}}
\end{eqnarray}
On the other hand,
\begin{eqnarray}
& & \sup_{y'}
|\tilde{J}_c(y')|\int_{\R}e^{-\beta(z)|y-y'|}|J'_c(y)|\, \D y\leq
\|\tilde{J}_c\|_{\infty}\,
\sup_{y'\in[-y_0,y_0]}\int_{y_0+F^{-\tau}}^{y_0+F^{-\tau}+1}e^{-\beta(z)|y-
y'|}\, \D y \nonumber \\
& & \leq \|\tilde{J}_c\|_{\infty}\, e^{-\beta(z)F^{-\tau}}
\end{eqnarray}
and similarly for the terms with $J''_c(y)$. When $|x-x'|<1$ we
proceed in a similar way as for the case $i=3$ and we get the
desired result.
\par Thus we can conclude that
\begin{equation} \label{ynorm}
\|[p_y^2,J_1(ib)]R_1(z;ib)\tilde{J}_1(ib)\|\leq {\cal C}\,
\beta(z)^{-\sigma}F^{-{\cal C}}e^{-\frac{\beta(z)}{F^{\tau}}}
\end{equation}
Finally,
$$
\|K_1(z;ib)\| \leq {\cal C}F^{-{\cal C}}
\beta(z)^{-\sigma}\left(e^{-\frac{\beta(z)}{F^{\tau}}}+e^{-\frac{{\cal
C}}{F^{2(1-\varepsilon)}}}\right)
$$

 \noindent The upper bound on the
term $\|K_5(z;ib)\|$ is found in the same way.

\subsubsection*{Norm of $K_2(z;ib)$}

The operator $K_2(z;ib)$ includes the resolvent $R_2(z;ib)$, which
can be evaluated with respect to $R_1(z;ib)$
\begin{equation}\label{reseq}
R_2(z;ib)=R_1(z;ib)-
R_1(z;ib)[F(x+ib)(\chi_A^c+h_F^c(ib)\chi_A)+V(ib)]R_2(z;ib)
\end{equation}
Obviously, the first term coming from (\ref{reseq}) is to be
treated in the same way as above. The second term
$R_1(z;ib)[\cdots]R_2(z;ib)$ is estimated using
$$
\|[H_L,J_2(ib)]R_1(z;ib)[\cdots]R_2(z;ib)\tilde{J}_2(ib)\|\leq
\|[H_L,J_2(ib)]R_1(z;ib)[\cdots]\|\|R_2(z;ib)\|\|\tilde{J}_2(ib)\|
$$
Now, $\|\tilde{J}_2(ib)\|$ is bounded and for $\|R_2(z;ib)\|$ we
use the result of Lemma \ref{R_2}. It then remains to estimate
\be\label{reseq1}
\|[H_L,J_2(ib)]R_1(z;ib)[F(x+ib)(\chi_A^c+h_F^c(ib)\chi_A)+V(ib)]\|
\ee

Before we give the estimation of the different contribution to
\eqref{reseq1}, we remind that
\begin{eqnarray}
|J'_0(x+ib)| &\leq& {\cal C}\, F^{-1} \left\{e^{-2\gamma|x-x_1|} +
e^{-2\gamma|x+x_1|}\right\} \\
|J''_0(x+ib)|&\leq& {\cal C}\, F^{-2} \left\{e^{-2\gamma|x-x_1|} +
e^{-2\gamma|x+x_1|}\right\},
\end{eqnarray}
where we have used the similar bounds as in \eqref{diff1}. In the
estimations we will separate the two contributions
coming from $\bar{J}_+$ and $\bar{J}_-$.\\

Let us now look at the contribution to ({\ref{reseq1}) which
includes the potential $V(ib)$. We again begin with the
Hilbert-Schmidt norm (case $|x-x'|\geq 1$) of the terms in the
commutator involving the $x-$derivatives. After separation of
variables we can write ($n=1,2$)
\begin{eqnarray}
& & \|\pd_x^n \bar{J}_+(x+ib)J_c(y)
\pd_x^{(2-n)}R_1(z;ib)V(ib)\|_{HS}^2
\nonumber\\
& & \leq {\cal C}\, F^{-2\tau}\beta(z)^{-\sigma}\, \int_{\R}
|\pd_x^{n}\bar{J}_+(x+ib)|^2 \D x\int_{\R}
e^{-\omega(x-x')^2}\, |V(x'+ib,\hat{y})|^2\D x'  \nonumber \\
& & \leq {\cal C}\, F^{-2-2\tau}\beta(z)^{-\sigma} \int_{\R}
e^{-4\gamma|x-x_1|}\left
  [\int_{|x'|\leq a_0}e^{-\omega(x-x')^2}\, \D x'
  +\int_{|x'|>a_0}e^{-\omega(x-x')^2}e^{-\nu x^{\prime 2}}\, \D x'\right
]\D x \nonumber \\
& & \leq {\cal C}\, F^{-2-2\tau}\beta(z)^{-\sigma} \int_{\R}
e^{-4\gamma|x-x_1|}\left [g(x,a_0)+\sqrt{\frac{\pi}{\omega+\nu}}\,
e^{-\frac{\omega\nu}{\omega+\nu}\, x^2}\right ]\D x
\end{eqnarray}
where we have defined
$$
g(x,a_0):=\int_{|x'|\leq a_0}e^{-\omega(x-x')^2}\, \D x'
$$
Now we can apply the same argument as in (\ref{3int}) and repeat
it for $\|\pd_x^n \bar{J}_-(x+ib)J_c(y)
\pd_x^{(2-n)}R_1(z;ib)V(ib)\|_{HS}^2$ to arrive at
\begin{equation} \label{Vgauss1}
\left \|[(p_x+By)^2,J_2(ib)]R_1(z;ib)V(ib)\right \|^2_{HS}\leq
{\cal C}\, \beta(z)^{-\sigma}F^{-{\cal C}}\, e^{-{\cal C}\,
F^{-2(1-\varepsilon)}}
\end{equation}
For $|x-x'|<1$ we proceed like in \eqref{petit1} evaluating
separately the contributions coming from $\bar{J}_+$ and
$\bar{J}_-$. For example, for $\pd_x^n \bar{J}_+(x+ib)J_c(y)
\pd_x^{(2-n)}R_1(z;ib)V(ib)$ we get an upper bound of the form
\bea & &\sup_{{\bf x}}
\sup_{x':|x-x'|<1,y'}|\partial_x^n\bar{J}_+'(x)V(x'+ib,y')|
\int_{\R} \D y' \int_{x':|x-x'|<1} \D x' |\partial_x^{2-n}G_1({\bf
x},{\bf x}';z)|\nonumber \\
&\leq& {\cal C}\beta(z)^{-\sigma}
\sup_{x,x':|x-x'|<1,y'}|\partial_x^n\bar{J}_+'(x)V(x'+ib,y')|\leq
{\cal C}\beta(z)^{-\sigma}F^{-{\cal C}} e^{-{\cal
C}F^{-2(1-\varepsilon)}}
 \label{Vgauss2}\eea
The last term in the commutator \eqref{reseq1} which includes
$V(ib)$ is the following
$$
[p_y^2,J_2(ib)]R_1(z;ib)V(ib)
$$
For $|x-x'|\geq 1$, since both
$$
J_0(x+ib)\int_{\R}e^{-\omega(x-x')^2}\D x', \quad
\int_{\R}e^{-\omega(x-x')^2}J_0(x'+ib)\D x'
$$
are bounded as functions of $x$, we apply again (\ref{maxnorm}) to
find out that
\begin{eqnarray}
& & \sup_y |J'_c(y)| V_0\int_{-a_1}^{a_1} e^{-\beta(z)|y-y'|} \D
y' \leq \|J'_c\|_{\infty}V_0\, 2a_1\, \sup_{y\in\supp
J'_c}\sup_{y'\in [-a_1,a_1]}e^{-\beta(z)|y-y'|}
\nonumber \\
& & \leq \|J'_c\|_{\infty}\, 2a_1 V_0\, e^{-\beta(z)F^{-\tau}}
\end{eqnarray}
and similarly the other way around
$$
\sup_{y'} |V(x'+ib,y')|\int_{y_0+F^{-\tau}}^{y_0+F^{-\tau}+1}
e^{-\beta(z)|y-y'|} |J'_c(y)|\D y  \leq  V_0 \|J'_c\|_{\infty}\,
e^{-\beta(z)F^{-\tau}}
$$
For $|x-x'|<1$ we proceed as for $i=3$. Summing all the above
given inequalities we obtain
\begin{equation}\label{Vib}
\left \|[H_L,J_2(ib)]R_1(z;ib)V(ib)\right \| \leq {\cal
C}\,\beta(z)^{-\sigma} F^{-{\cal C}}
\left(e^{-\frac{\beta(z)}{F^\tau}} +e^{-\frac{{\cal C}}{
F^{2(1-\varepsilon)}}} \right)
\end{equation}

\begin{rem}\label{condonV}
Note that the hypothesis on the Gaussian-like decay of $V$ w.r.t.
$x$ is necessary in order to obtain \eqref{Vib} as one can see
from \eqref{Vgauss1} and \eqref{Vgauss2}.
\end{rem}

Next we analyse those terms of (\ref{reseq1}), which include the
potential $F(x+ib)h_F^c(ib)\chi_A$. We start again with the case
$|x-x'|\geq 1$ looking at the Hilbert-Schmidt norm of
\begin{equation} \label{W_F}
[(p_x+By)^2,J_2(ib)]R_1(z;ib)F(x+ib)h_F^c(ib)\chi_A
\end{equation}
Note that since we have the same upper bounds on $J'_c(x+ib),\,
J''_c(x+ib)$ and also on $R_1(z;ib),\, \pd_x R_1(z;ib)$, all terms
in (\ref{W_F}) can be estimated in the same way. As for the
previous term we separate the contributions of $\bar{J}_\pm$,
moreover $h_F^c=1-h_F=h_++h_-$ with
$h_\pm(x)=\tfrac{1}{2}\left[1\mp\tanh (\gamma_F(x\pm
\bar{x}))\right]$, and thus we separate also the contributions of
$h_{+}$ and $h_{-}$. We are left with four terms, each of them is
estimated as follows ($n=1,2$):
\begin{eqnarray}
& & \|\pd_x^n \bar{J}_+(x+ib)J_c(y)
\pd_x^{(2-n)}R_1(z;ib)F(x+ib)h_-
(ib)\chi_A\|^2_{HS}  \nonumber\\
& & \leq {\cal C}\, \beta(z)^{-\sigma}F^{-{\cal C}}\, \int_{\R}
|\pd_x^{n}\bar{J}_+(x+ib)|^2 \D x\int_{\R} e^{-\omega(x-x')^2}\,
|F(x'+ib)h_-(x'+ib)|^2\D x'  \nonumber \\
& & \leq {\cal C}\, \beta(z)^{-\sigma}F^{-{\cal
C}}\int_{\R}e^{-4\gamma |x-x_1|}\D
x\int_{\R}e^{-\omega(x-\bar{x}-t)^2}|t+\bar{x}+ib|^2\, \frac{\D
t}{1+e^{-4\gamma t}}
\end{eqnarray}
recalling that the integration w.r.t. $y,y'$ gives again the
factor of order $F^{-2\tau}$. To evaluate the integral with
respect to $t$ we write
\begin{eqnarray}
& & \int_{\R}e^{-\omega(x-\bar{x}-t)^2}|t+\bar{x}+ib|^2\, \frac{\D
t}{1+e^{-4\gamma t}}  \\
& & \leq \int_{-\infty}^0e^{-\omega(x-\bar{x}-t)^2+4\gamma
t}(2t^2+2\bar{x}^2+b^2)\, \D t +
\int_0^{\infty}e^{-\omega(x-\bar{x}-t)^2}(2t^2+2\bar{x}^2+b^2)\,
\D t \nonumber
\end{eqnarray}
and use the following general result which can be found in
\cite[p. 1064]{GR},
\begin{equation} \label{integral}
\int_{0}^{\infty}t^{\mu-1}e^{-b t^2-c t}\, \D t =(2b)^{-
\mu/2}\Gamma(\mu)\exp(c^2/8b) D_{-\mu}(c/\sqrt{2b})
\end{equation}
Here $D_{-\mu}(\cdot)$ is the parabolic cylinder function of order
$-\mu$. Its asymptotic behaviour is given by \cite[p.1065]{GR}
\be\label{asympt}
\begin{array}{lclc}
D_{p}(z) & \simeq & e^{-z^2/4}z^{p}(1-\mathcal{O}(z^{-2})), & z\ra\infty \\
D_{p}(z) & \simeq & e^{z^2/4}z^{-p-1}(1+\mathcal{O}(z^{-2})), &
z\ra-\infty
\end{array}
\ee The asymptotic behaviour allows us to apply once more the
argument used in (\ref{3int}). We can thus claim that
$$
\left \|[(p_x+By)^2,J_2(ib)]R_1(z;ib)F(x+ib)h_F^c(ib)\chi_A\right
\|^2_{HS}\leq {\cal C}\,\beta(z)^{-\sigma} F^{-{\cal C}}\,
e^{-{\cal C}\, F^{-2(1-\varepsilon)}}
$$
Also for the case $|x-x'|<1$ all the terms are treated
analogously. For example for
$\partial_x^n\bar{J}_+(ib)J_c\partial_x^{2-n}
R_1(z;ib)F(x+ib)h_-(ib)\chi_A$ we have
\begin{eqnarray}\label{ouf}
& &\sup_{{\bf x}} \int_{\R} \D y' \int_{|x- x'|<1} \D x'
|\partial_x^n\bar{J}_+(x+ib)J_c(y)||\partial_x^{2-n}G_1({\bf
x},{\bf
x}';z)|F|x'+ib||h_-(x'+ib)\chi_A(y')|\nonumber \\
&\leq&\sup_{{\bf x}}
\sup_{x':|x-x'|<1}|\partial_x^n\bar{J}_+(x+ib)h_-(x'+ib)|^{1/2}\times
\\
&\times& \int_{\R} \D y' \int_{|x-x'|<1} \D x'
|\partial_x^{2-n}G_1({\bf x},{\bf
x}';z)||\partial_x^n\bar{J}_+(x)|^{1/2}F|x'+ib| \leq {\cal
C}\beta(z)^{-\sigma}F^{-{\cal C}}e^{-{\cal
C}F^{-2(1-\varepsilon)}} \nonumber
 \label{petit2}
\eea where we used the fact that $|x'|\leq |x|+1$ and
$|\partial_x^n\bar{J}_+(x)|^{1/2}|x|\leq {\cal C}F^{-(1-\varepsilon)}$.\\
We are now left with the last term in the commutator:
\begin{eqnarray}
& &
[p_y^2,J_2(ib)]R_1(z;ib)F(x+ib)h_F^c(ib)\chi_A=-2J_0(x+ib)J'_c(y)\pd_y
R_1(z;ib)\times \nonumber \\
& & \times F(x+ib)h_F^c(ib)\chi_A -J_0(x+ib)J''_c(y)
R_1(z;ib)F(x+ib)h_F^c(ib)\chi_A
\end{eqnarray}

When $|x-x'|\geq 1$ the Hilbert-Schmidt norm of these terms can
estimated separately for $h_\pm$. We do that for $h_-$, for the
term coming from
$h_+$ a similar argument holds.\\
For $h_-$ the Hilbert-Schmidt norm is bounded above by a constant
times $\beta(z)^{-\sigma}F^{-\tau}$ (coming from the integration
w.r.t. $y$ and $y'$) times
\begin{eqnarray} \label{appenc}
& &  \int_{\R}\D x|J_0(x+ib)|^2\,
\int_{\R}e^{-\omega(x-x')^2}|x'|^2\frac{\D x'}
{1+e^{-4\gamma(x'-\bar{x})}} \nonumber \\
& & \leq  \int_{\R}\D x|J_0(x+ib)|^2\,
\int_{\R}e^{-\omega(x-\bar{x}-t)^2}(2t^2+2\bar{x}^2)\, \frac{\D
t}{1+e^{-4\gamma t}}
\end{eqnarray}
The last integral can be again evaluated through (\ref{integral})
and \eqref{asympt} and estimated up to a constant from above by
\begin{equation}\label{appenC}
 F^{-{\cal C}}\, e^{-{\cal C}\, F^{-2(1-\varepsilon)}},
\end{equation}
To control the first term in (\ref{appenc}), which is proportional
to $t^2$, we proceed in the same way as in (\ref{division}) to
write \bea \label{cylinder} & &
\int_{\R}e^{-\omega(x-\bar{x}-t)^2}\, t^2\, \frac{\D
t}{1+e^{-4\gamma t}} \leq {\cal C}\, e^{-\omega(x-\bar{x})^2}\Bigg
[e^{\frac{\omega(x-\bar{x})^2}{2}} D_{-3}(\sqrt{2\omega}\,
(\bar{x}-x))
\nonumber \\
& + & e^{\frac{(2\omega(x-\bar{x})+4\gamma)^2}{8\omega}}
D_{-3}\left(\frac{2\omega(x-\bar{x})+4\gamma}{\sqrt{2\omega}}\right
)\Bigg ] \eea We will split (\ref{appenc}) in three parts: \be
\label{threeparts} (-\infty,x_1 +\delta],\quad [x_1 +\delta,
\bar{x}],\quad [\bar{x},\infty) \ee where $\delta=\delta_0\,
F^{-(1-\varepsilon)}$ and $(x_1 +\delta)<\bar{x}$. For the first
part we get \bea & & \int_{\bar{x}}^{\infty}\D x
e^{-4\gamma(x-x_1)}\, e^{-\omega(x-\bar{x})^2/2}
D_{-3}(\sqrt{2\omega}(\bar{x}-x)) \nonumber\\
& & \leq e^{-4\gamma(\bar{x}-x_1)}\int_0^{\infty} e^{-4\gamma
t-\omega t^2/2} D_{-3}(-\sqrt{2\omega}\, t)\D t  \leq {\cal C}\,
e^{-4\gamma(\bar{x}-x_1)}
\end{eqnarray}
since $ e^{-4\gamma t-\omega t^2/2} D_{-3}(-\sqrt{2\omega}\, t)$
is clearly $L_1([0,\infty))$, see (\ref{asympt}). The second part
can be estimated as follows \bea\label{2nd} & & \int_{x_1
+\delta}^{\bar{x}}\D x  e^{-4\gamma(x-x_1)}\,
e^{-\omega(x-\bar{x})^2/2}
D_{-3}(\sqrt{2\omega}(\bar{x}-x)) \D x \nonumber\\
& & \leq e^{-4\gamma\delta}  \int_{x_1 +\delta}^{\bar{x}}
e^{-\omega(x-\bar{x})^2/2}
D_{-3}(\sqrt{2\omega}(\bar{x}-x))\D x \nonumber \\
& & \leq  e^{-4\gamma\delta}(\bar{x}-x_1 -\delta)\sup_{x\in[x_1
+\delta, \bar{x}]} D_{-3}(\sqrt{2\omega}(\bar{x}-x))  \leq {\cal
C}\, F^{-(1-\varepsilon)}\, e^{-4\gamma\delta},\quad F\ra 0
\end{eqnarray}
Finally, the third part is bounded above by \bea & &
\int_{-\infty}^{x_1 +\delta} e^{-\omega(x-\bar{x})^2/2}
D_{-3}(\sqrt{2\omega}(\bar{x}-x))\D x \leq
e^{-\omega\bar{x}^2/2}\int_{-\infty}^0
D_{-3}(\sqrt{2\omega}(\bar{x}-x))\D
x \nonumber \\
& & + e^{-\omega(\bar{x}-x_1 -\delta)^2/2} \int_0^{x_1 +\delta}
D_{-
3}(\sqrt{2\omega}(\bar{x}-x))\D x \nonumber \\
& & \leq {\cal C}\, e^{-\omega(\bar{x}-x_1 -\delta)^2/2}, \quad
F\ra 0
\end{eqnarray}
where we have employed the asymptotic expansion (\ref{asympt}).
\par The estimate of the second part of (\ref{cylinder}), which
contains the function \be
 D_{-3}\left(\frac{2\omega(x-\bar{x})+4\gamma}{\sqrt{2\omega}}\right )
\end{equation}
is a bit more subtle. After dividing the integration again in
three parts according to (\ref{threeparts}) and substituting \be
t:= \frac{2\omega(x-\bar{x})+4\gamma}{\sqrt{2\omega}} \ee one gets
\bea & & \int_{\bar{x}}^{\infty}\D xe^{-4\gamma(x-x_1)}
e^{-\omega(x-\bar{x})^2}
e^{\frac{(2\omega(x-\bar{x})+4\gamma)^2}{8\omega}}
D_{-3}\left(\frac{2\omega(x-\bar{x})+4\gamma}{\sqrt{2\omega}}\right
)
\nonumber \\
& & \leq
e^{-4\gamma(\bar{x}-x_1)}\int_{4\gamma/\sqrt{2\omega}}^{\infty}\exp\left
[
  -\frac{t^2}{4}+\frac{2\sqrt{2}\, \gamma }{\sqrt{\omega}}\,
  t-\frac{4\gamma^2}{\omega}\right ]
D_{-3}(t) \sqrt{2\omega}\D t  \nonumber \\
& &\leq {\cal C}\, e^{-{\cal C}\, F^{-2(1-\varepsilon)}}, \quad
F\ra 0
\end{eqnarray}
provided \be \label{gammacond} \omega (\bar{x}-x_1)>\gamma \ee
this can be seen taking the maximum of the exponential function in
the integral and the fact that $D_{-3}(t)\in L_1([0,\infty))$.

 For
$x\in(-\infty, x_1 +\delta]$ we have similarly \bea & &
\int_{-\infty}^{x_1 +\delta}\D x
e^{-\omega(x-\bar{x})^2}e^{\frac{(2\omega(x-\bar{x})+4\gamma)^2}{8\omega}}
D_{-3}\left(\frac{2\omega(x-\bar{x})+4\gamma}{\sqrt{2\omega}}\right
)
\nonumber \\
& & \leq \int_{-\infty}^{\frac{2\omega(x_1
    +\delta-\bar{x})+4\gamma}{\sqrt{2\omega}}} \, \exp\left [
  -\frac{t^2}{4}+\frac{2\sqrt{2}\, \gamma }{\sqrt{\omega}}\,
  t-\frac{4\gamma^2}{\omega}\right ]\, D_{-3}(t)  \sqrt{2\omega}\D t
\eea Since \be\label{L_1} \exp\left [
-\frac{t^2}{4}+\frac{2\sqrt{2}\, \gamma }{\sqrt{\omega}}\, t
\right ]\, D_{-3}(t)\in L_1((-\infty,0]) \ee it suffices to
estimate the integral for positive values of $t$. In this case we
use the fact that $$D_{-3}(z)e^{\xi z^2/4} \in L_1([0,\infty)),
$$for any $\xi<1$. Then \bea & & \int_0^{\frac{2\omega(x_1
+\delta-\bar{x})+4\gamma}{\sqrt{2\omega}}} \, \exp\left [
-\frac{t^2(1+\xi)}{4}+\frac{2\sqrt{2}\, \gamma }{\sqrt{\omega}}\,
  t-\frac{4\gamma^2}{\omega}\right ] e^{\xi t^2/4}\, D_{-3}(t)
\sqrt{2\omega} \D t \nonumber
\\
& & \leq {\cal C}\, e^{-{\cal C}\, F^{-2(1-\varepsilon)}}, \quad
F\ra 0 \eea whenever $$1>\xi> \frac{4\gamma^2-\omega^2(x_1
+\delta-\bar{x})^2}{4\gamma^2+\omega^2(x_1
+\delta-\bar{x})^2}=\frac{4\gamma_0^2-\omega^2(C_1
+\delta_0-\bar{C})^2}{4\gamma_0^2+\omega^2(C_1
+\delta_0-\bar{C})^2} $$We are thus left with \bea & & \int_{x_1
+\delta}^{\bar{x}}\D x \, e^{-4\gamma(x-x_1)}
e^{-\omega(x-\bar{x})^2}e^{\frac{(2\omega(x-\bar{x})+4\gamma)^2}{8\omega}}
D_{-3}\left(\frac{2\omega(x-\bar{x})+4\gamma}{\sqrt{2\omega}}\right
)
\nonumber \\
& & \leq e^{-4\gamma\delta} \int_{\frac{2\omega(x_1
+\delta-\bar{x})+4\gamma}{\sqrt{2\omega}}}
^{\frac{4\gamma}{\sqrt{2\omega}}}\exp\left [
-\frac{t^2}{4}+\frac{2\sqrt{2}\,
 \gamma }{\sqrt{\omega}}\, t -\frac{4\gamma^2}{\omega}\right ]\, D_{-3}(t)
 \sqrt{2\omega}\D t
\eea Due to (\ref{L_1}) it is enough to show that \be
 \int_0^{\frac{4\gamma}{\sqrt{2\omega}}}\exp\left [
 -\frac{t^2}{4}+\frac{2\sqrt{2}\, \gamma }{\sqrt{\omega}}\, t
 -\frac{4\gamma^2}{\omega}\right ]\, D_{-3}(t)  \sqrt{2\omega}\D t \leq
{\cal C}\,
 F^{-(1-\varepsilon)}\,
 \ee
This is however easily seen since \be
-\frac{t^2}{2}+\frac{2\sqrt{2}\, \gamma }{\sqrt{\omega}}\, t
-\frac{4\gamma^2}{\omega}\leq 0, \quad \forall\, t\in \left
  [0,\frac{4\gamma}{\sqrt{2\omega}}\right ]
\ee and $$\sup_{t\in[0,\frac{4\gamma}{\sqrt{2\omega}}]}
e^{t^2/4}D_{-3}(t)\leq \sup_{t\in[0,\infty)}
e^{t^2/4}D_{-3}(t)\leq {\cal C} $$To conclude we remark that the
second term of (\ref{appenc}), which leads to \bea & &
\int_{\R}e^{-\omega(x-\bar{x}-t)^2}\, \bar{x}^2\, \frac{\D
t}{1+e^{-4\gamma t}}  \leq {\cal C}\, F^{-2(1-\varepsilon)}\,
e^{-\omega(x-\bar{x})^2}\Bigg [e^{\frac{\omega(x-\bar{x})^2}{2}}
D_{-1}(\sqrt{2\omega}\, (\bar{x}-x)) \nonumber \\
& + & e^{\frac{(2\omega(x-\bar{x})+4\gamma)^2}{8\omega}}
D_{-1}\left(\frac{2\omega(x-\bar{x})+4\gamma}{\sqrt{2\omega}}\right
)\Bigg ], \eea can be control in the same way, because the
asymptotic behaviour (\ref{asympt}) is again governed by $\exp[\pm
t^2/4]$.\\
Finally, for the case $|x-x'|<1$ we follows the same method as in
\eqref{ouf} where the decay come from the ``infinitesimally
small'' overlap of $h_F^c$ with $J_0$ the latter also ``localise''
$|x'|$, i.e. $|J_0(x+ib)|^{1/2}|(x'+ib)|\leq{\cal C}
F^{-(1-\varepsilon)}$. Summing up all the contributions we have
\begin{equation}\label{hj1}
\left \|[H_L,J_2(ib)]R_1(z;ib)F(x+ib)h_F^c(ib)\chi_A\right \| \leq
{\cal C}\,\beta(z)^{-\sigma} F^{-{\cal C}} e^{-\frac{\cal C}{
F^{2(1-\varepsilon)}}}
\end{equation}
$ $\\
Let us next analyse the last term of (\ref{reseq1}), which
includes the potential $F(x+ib)\chi^c_A$. When $|x-x'|\geq 1$, for
the terms in the commutator involving the $x-$derivatives, the
integration w.r.t. $x$ and $x'$ in the Hilbert-Schmidt norm gives
a constant proportional to $F^{-2(1-\varepsilon)}$. We then obtain
the estimate on the Hilbert-Schmidt norm
\begin{eqnarray}
&&  \|\pd_x^n J_2(ib)
\pd_x^{(2-n)}R_1(z;ib)F(x+ib)\chi^c_A\|^2_{HS}\nonumber
  \\
&& \leq {\cal C}\, \beta(z)^{-\sigma}F^{-{\cal C}}\,
 \int_{-y_1}^{y_1}\D y \int_{|y'|\geq y_1+F^{-\tau}} e^{-2\beta(z)|y-y'|}\,
\D y'  \nonumber \\
 && \leq {\cal C}\,  \beta(z)^{-\sigma}F^{-{\cal C}}\,
e^{-\frac{\beta(z)}{F^\tau}}\int_{-\infty}^{\infty}e^{-\beta(z)\,
|y-y'|}\D y'  \nonumber\\
&& \leq {\cal C}\,  \beta(z)^{-\sigma}F^{-{\cal C}}\,
e^{-\frac{\beta(z)}{F^{\tau}}}
\end{eqnarray}
When $|x-x'|<1$ the $x-$derivative ``localises'' the term
$|x'+ib|$ and the decay comes from the decay of the Green function
along $y$ as for the case $i=3$.\\
 For the term of the commutator
which corresponds to
$$
\pd_y^n J_2(ib) \pd_y^{(2-n)}R_1(z;ib)F(x+ib)\chi^c_A,\quad
|x-x'|\geq 1, \quad n=1,2
$$
we recall (\ref{maxnorm}) to find out that
\begin{eqnarray}
& & \sup_{{\bf x}}\int_{\R^2}|J_0(x+ib)\pd_y^nJ_c(y)
\pd_y^{(2-n)}G_1({\bf x},{\bf x}';z)F(x'+ib)\chi^c_A(y')|\D {\bf
x}'
\nonumber \\
& & \leq \frac{\cal C}{F^{1-\varepsilon}}\,
\beta(z)^{-\sigma}\sup_{y\in\supp \pd_y^nJ_c}\, \int_{|y'|\geq
y_1+F^{-\tau}}
e^{-\beta(z)|y-y'|}\D y' \nonumber \\
& & \leq {\cal C}\, \beta(z)^{-\sigma}F^{-{\cal C}}\,
e^{-\frac{\beta(z)}{F^{\tau}}}
\end{eqnarray}
and similarly the other way around. Finally at short distances the
same argument as in the previous case holds. Therefore
\begin{equation}\label{hj2}
\left \|[H_L,J_2(ib)]R_1(z;ib)F(x+ib)\chi^c_A\right \| \leq {\cal
C}\,\beta(z)^{-\sigma} F^{-{\cal C}} e^{-\frac{\beta(z)}{F^\tau}}
\end{equation}

Taking into account all the estimates \eqref{Vib}, \eqref{hj1},
\eqref{hj2} made above, we can claim that for $F$ small enough
\begin{equation} \label{essential}
\left \|K_2(z;ib)\right \| \leq {\cal C}\, F^{-{\cal C}}\,
\beta(z)^{-\sigma(z)} \left(e^{-\frac{\beta(z)}{F^\tau}}
+e^{-\frac{{\cal C}}{ F^{2(1-\varepsilon)}}} \right) \left
(1+\|R_2(z;ib)\| \right )
\end{equation}

Inequality \eqref{essential} plays an essential role in our
estimates, because it tells us how close we can get to the
spectrum of $H_2(F,ib)=H_2(F)$ and $H_1(F,ib)$ while keeping the
resolvent of $H(F,ib)$ bounded.

\section*{Acknowledgements}
We wish to thank P.A.Martin and N.Macris for suggesting to us the
presented problem and for many stimulating and encouraging
discussions throughout the project. Numerous comments of P.Exner
are also gratefully acknowledged. H.K. would like to thank his
hosts at Institute for Theoretical Physics, EPF Lausanne for a
warm hospitality extended to him. C.F. thanks the Math. department
at Stuttgart University for hospitality, where part of the present
work was done.  The work of C.F. was supported by the Fonds
National Suisse de la Recherche Scientifique No. 20-55694.98.

\end{document}